\documentclass[aps,pra,twocolumn,superscriptaddress,showpacs,amsmath,amstex,amssymb,citeautoscript]{revtex4-1}

\usepackage[T1]{fontenc}
\usepackage[utf8]{inputenc}
\usepackage{lipsum}
\usepackage{amsmath}
\usepackage{amssymb}
\usepackage{bbm}
\usepackage{braket}
\usepackage{xcolor}
\usepackage{pifont}
\usepackage{ragged2e}
\usepackage{blindtext}
\usepackage[mathscr]{euscript}
\usepackage[shortlabels]{enumitem}
\usepackage[justification=justified]{subcaption}
\usepackage{graphicx}
\usepackage{lipsum}
\allowdisplaybreaks
\usepackage{float}
\usepackage{graphicx}
\usepackage{dsfont}
\usepackage{comment}
\usepackage[normalem]{ulem}
\usepackage[colorlinks=true]{hyperref}  
\hypersetup{
    bookmarks=true,         
    unicode=false,          
    pdftoolbar=true,        
    pdfmenubar=true,        
    pdffitwindow=false,     
    pdfstartview={FitH},    
    pdftitle={Correlated states of twisted mono-bilayer graphene
with proximity-induced spin orbit coupling},    
    pdfauthor={Park, Scheurer, Classen},     
    pdfsubject={},   
    pdfcreator={},   
    pdfproducer={}, 
    pdfkeywords={} {} {}, 
    pdfnewwindow=true,      
    colorlinks=true,       
    linkcolor=blue, 
    citecolor=blue,        
    filecolor=magenta,      
    urlcolor=blue           
}

\newcommand{\equref}[1]{Eq.~(\ref{#1})}

\newcommand{\secref}[1]{Sec.~\ref{#1}}
\newcommand{\figref}[1]{Fig.~\ref{#1}}

\newcommand{\appref}[1]{Appendix~\ref{#1}}

\newcommand{\pdagger}{{\phantom{\dagger}}}

\renewcommand{\vec}[1]{\boldsymbol{#1}}

\definecolor{wrongultramarine}{rgb}{1,0.5,0}

\begin{document}

\title{Tuning correlated states of twisted mono-bilayer graphene \\
with proximity-induced spin-orbit coupling} 

\affiliation{Max Planck Institute for Solid State Research, D-70569 Stuttgart, Germany}
\affiliation{School of Natural Sciences, Technische Universit\"at M\"unchen, D-85748 Garching, Germany}

\author{Jeyong Park}
\affiliation{Max Planck Institute for Solid State Research, D-70569 Stuttgart, Germany}
\affiliation{School of Natural Sciences, Technische Universit\"at M\"unchen, D-85748 Garching, Germany}
\author{Mingdi Luo}
\affiliation{Max Planck Institute for Solid State Research, D-70569 Stuttgart, Germany}
\author{Louk Rademaker}
\affiliation{Department of Theoretical Physics, University of Geneva, CH-1211 Geneva, Switzerland}
\affiliation{Institute-Lorentz for Theoretical Physics, Leiden University,
PO Box 9506, Leiden, NL-2300, The Netherlands}
\author{Jurgen Smet}
\affiliation{Max Planck Institute for Solid State Research, D-70569 Stuttgart, Germany}
\author{Mathias S.~Scheurer}
\affiliation{Institute for Theoretical Physics III, University of Stuttgart, 70550 Stuttgart, Germany}
\author{Laura Classen}
\affiliation{Max Planck Institute for Solid State Research, D-70569 Stuttgart, Germany}
\affiliation{School of Natural Sciences, Technische Universit\"at M\"unchen, D-85748 Garching, Germany}

\begin{abstract}
We study the correlated ground states of twisted mono-bilayer graphene  with and without proximity-induced spin-orbit coupling (SOC) from a transition-metal dichalcogenide layer placed on top. 
We perform self-consistent Hartree-Fock calculations that allow the variational space to include multi-$Q$ translational symmetry broken states for all integer and half-integer fillings of the conduction bands, where signatures of correlated, topological states have been reported experimentally. We find interaction-induced insulators that retain moir\'e translational symmetry at integer fillings, but that break this symmetry at half-integer fillings. 
We argue that translational symmetry breaking arises from half-filled polarized bands, even when SOC is present. Yet, we find that small SOC can already crucially affect the spin nature of correlated states. Generally, Ising SOC favors out-of-plane spin polarization and spin-valley locking, while Rashba SOC favors in-plane spin order. If only one of these two terms is present, we find that, depending on the type of SOC, it drives a transition from a tetrahedal antiferromagnet to either a coplanar, non-coplanar, or collinear spin-density wave state for half-integer fillings. The frustration associated with the simultaneous presence of both types of SOC can induce chiral, non-coplanar order in parameter ranges where the ground state in the absence of SOC is collinear.
\end{abstract}

\maketitle

\section{Introduction}
Twisted moir\'e materials have recently become a versatile platform for studying strongly correlated electron behavior. Their properties can be controlled by various tuning knobs such as the twist angle between the layers, external fields, or substrate variation. These parameters facilitate the design of flat or narrow band dispersions that may possess non-trivial topology and can give rise to strong interaction and correlation effects such as the appearance of correlated insulators, superconductivity, as well as anomalous quantum Hall states 
\cite{jiang2019charge, cao2018unconventional,  park2021tunable, kim2022evidence, chen2021electrically,  xu2021tunable,polshyn2020electrical, he2021competing,polshyn2022topological,wang2025moir,liu2025diverse,dong2025observation, wang2026family, su2025moire, zhang2025entangled}.  Theoretically, it was also suggested that moir\'e materials can mimic different model systems ranging, for example, from 1D  helical network models, over Hubbard models, to heavy-fermion models \cite{xie2023phase,chichinadze2020nematic, park2023network, classen2019competing, efimkin2018helical,po2018origin, song2022magic, PhysRevLett.127.026401, ingham2025moir, zhang2019twisted, hawashin2025relativistic, shi2022heavy, yu2023magic, huang2024evolution, huang2025angle}. 
Given their extreme versatility and tunability, moir\'e systems might therefore even be thought of as a novel route for quantum simulation \cite{kennes2021moire}.

\begin{figure}[b]
 \centering
 \includegraphics[scale=0.5]{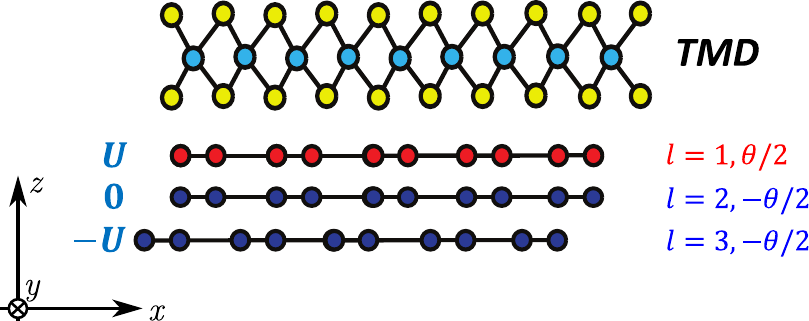}
	\caption{{\bf Schematic cross-section of SOC-TMBG:} a van der Waals heterostructure composed of a graphene monolayer ($l = 1$) and Bernal-stacked graphene bilayer ($l=2$ and $l=3$) with a relative twist angle of $\theta$, as well as a TMD layer.
    \justifying
    The TMD layer, such as  WSe${}_2$, is located on top of the graphene monolayer. We study both the case with and without the TMD layer. A displacement field is applied along the $\vec{z}$ direction and the associated electrostatic potential is designated as $U,0,-U$ for layer $l=1,2,3$, respectively. 
	}
	\label{Fig0:Setup}
\end{figure}

An interesting member of the graphene-based moir\'e materials family is twisted mono-bilayer graphene (TMBG) which goes beyond the high-symmetry configurations of alternately twisted few-layer graphene. Here, we theoretically investigate its correlated ground states with and without proximity-induced spin-orbit coupling (SOC) from a transition metal dichalcogenide (TMD) layer such as WSe$_2$, see Fig.~\ref{Fig0:Setup}. The single-particle band structure of TMBG can be manipulated via the displacement field. In particular, the minibands around the Fermi level can become narrow and topological with Chern numbers $C=2$ and $C=-1$ for a given valley \cite{rademaker2020topological, park2020gate, wang2024chern}, thereby creating a strongly correlated electron system prone to interaction effects. Indeed, in TMBG without a TMD layer but with an applied displacement field, experimental signatures of correlated insulators were recently reported at integer ($\nu=1,2,3$) and half-integer filling ($\nu=3/2,7/2$) of the conduction band~\cite{polshyn2020electrical, xu2021tunable, chen2021electrically, he2021competing,polshyn2022topological}. The appearance of these correlated insulators was accompanied by the quantum anomalous Hall effect with quantized values corresponding to either a Chern number of $C=2$ for integer fillings  $\nu=1$ and $\nu=3$ or a Chern number of $C=1$ for half-integer fillings. This behavior is consistent with interaction-induced valley polarization at integer filling and the emergence of a topological charge density wave state (CDW) \cite{polshyn2022topological} or a competing tetrahedral antiferromagnetic state \cite{wilhelm2023non} at 
half-integer fillings. It is natural to think of the states at the integer fillings $\nu = 1,2,3$ as polarized so that $\nu$ out of four originally spin- and valley-degenerate bands are filled. Similarly, at half-integer filling $\nu=3/2,7/2$ spontaneously breaking translation symmetry is consistent with splitting the Chern 2 band into two Chern 1 bands. However, it is an open question how these states evolve as a function of filling, because a complete phase diagram for all these fillings is still missing.

An interesting possibility is also to manipulate these correlated ground states by placing a proximitizing TMD layer on top of the TMBG. The TMD layer induces Ising and Rashba SOC, so that the spin degeneracy of the bands can be lifted. For other graphene-based moir\'e materials, it has been reported that adding a TMD layer can boost ferromagnetic or superconducting behavior \cite{PergeSOC,lin2022spin, scammell2023tunable, chou2024topological, tan2024topological, su2023superconductivity}. In TMBG, the effect of SOC is still unclear.

Therefore, we address here the influence of SOC on the correlated ground states of TMBG by performing self-consistent Hartree-Fock calculations for half-integer and integer fillings of the lowest conduction minibands. We compare these results with calculations obtained in the absence of proximity-induced SOC.  We focus on a finite displacement field. The reason is that, as stated above, previous experimental work has shown signatures of correlated Chern insulators in TMBG that is not covered by a TMD layer when a finite displacement field is applied~\cite{polshyn2020electrical, polshyn2022topological, xu2021tunable, chen2021electrically, he2021competing}. Moreover, under these conditions, single-particle band structure calculations reveal isolated conduction minibands with the narrowest width in the presence of a finite displacement field~\cite{polshyn2020electrical, polshyn2022topological, xu2021tunable, chen2021electrically, he2021competing,rademaker2020topological, park2020gate, wang2024chern}. We allow the HF ans\"atze to also capture translational symmetry broken states on the scale of the moir\'e lattice. 
Motivated by the aforementioned experimental and theoretical works, we use the three $M$ points in the moir\'e Brillouin zone (mBZ) as the wave vectors for the states with broken translational symmetry, i.e., our calculation can describe single-$Q$ spin density wave (SDW) and multi-$Q$ SDW states with a charge modulation in addition to a translation-invariant (TI) state with flavor symmetry breaking (such as spin or valley polarization).
For TMBG, we find interaction-induced insulators either due to TI polarized orders at integer fillings, or due to states with broken translational symmetry at half-integer fillings. Furthermore, we show that a multi-$Q$ state is generally favored over a single-$Q$ state. 

To systematically study how SOC affects the correlation physics, we perform HF calculations for different combinations of Ising and Rashba SOC parameters. 
We find that SOC can change the nature of the ground states. Generally, Ising SOC induces additional SVP, while Rashba SOC suppresses it. The occurrence of translational symmetry breaking remains robust with the addition of SOC for all considered half-integer fillings. In addition, translational symmetry broken states also become energetically closer to TI states for some integer fillings and some SOC parameters. However, SOC strongly affects the type of multi-$Q$ broken translational symmetry states. It influences the emergence of in-plane or out-of-plane spin order and changes the energetic competition between TAF, coplanar, non-coplanar, and collinear spin-density wave states. As such, it can be used to tune transitions between a non-coplanar, coplanar, and collinear order.
Lastly, we show that including the valence bands into our HF analysis does not qualitatively change our results. In addition, we see that band hybridization enhances the indirect band gap between the conduction and valence bands, especially near charge neutrality. 

This paper is organized as follows. In Sec.~\ref{Sec:ModelAndMethod} we introduce the continuum model for TMBG including proximity-induced SOC in the form of Ising and Rashba SOC. This yields the non-interacting band structure visualized in Fig.~\ref{Fig1:TMBGSOC}. The addition of Coulomb interaction and the Hartree-Fock approximation are described in Sec.~\ref{sec:HF}. The resulting phase diagram for translationally invariant and stripe phases is discussed in Sec.~\ref{Sec:HFconductionBands}. The more complicated multi-$Q$ states are discussed first using the Ginzburg-Landau theory in Sec.~\ref{GLSection}, followed by Hartree-Fock calculations in Sec.~\ref{Sec:HF3Q}. We conclude with an outlook in Sec.~\ref{Sec:Conclusion} on the possible experimental detection of the non-trivial phases we predict.

\section{Model and Method}
\label{Sec:ModelAndMethod}
\subsection{Single-particle Hamiltonian of SOC-TMBG}
To define the model we study in this work, let us start from the single-particle Hamiltonian for SOC-TMBG with the lattice geometry shown in \figref{Fig0:Setup}. For the TMBG part, we consider the case where a graphene monolayer is placed on top of an AB-stacked graphene bilayer with a relative twist angle of $\theta$. In this configuration, only the graphene monolayer experiences the proximity-induced SOC effect, since the TMD layer is located on top of TMBG. We also apply a perpendicular displacement field to the SOC-TMBG. This is modeled by applying an electrostatic potential of $U$ to the graphene monolayer, $0$ to the top layer of the bilayer and $-U$ to the bottom layer. To address the low-energy physics of this system, we consider a continuum model. In momentum space, it is written as
\begin{align}\begin{split}
&H^{\tau}(\vec{k},\vec{k}') = \\ &\,\, 
\begin{pmatrix} [H^{\tau}_{\textrm{MG}}(\vec{k}_{-\theta/2})+H^{\tau}_{\textrm{SOC}}]\delta_{\vec{k}',\vec{k}} & [T^{\tau}]_{\vec{k},\vec{k}'} \\
           [T^{\tau \dagger}]_{\vec{k},\vec{k}'} &H^{\tau}_{\textrm{BG}}(\vec{k}_{\theta/2})\delta_{\vec{k}',\vec{k}}
             \end{pmatrix}  ,
\end{split}\label{eq:Ham}\end{align}
with $\tau = \pm 1$ for valley $K$ and $K'$ respectively. Here, we use the rotated momenta $\vec{k}_{\theta} = R(\theta)\vec{k}$, where the matrix $R(\theta)$ describes a counterclockwise rotation of angle $\theta$. The single-particle Hamiltonian of the graphene monolayer and bilayer near around the $K$ $(\tau=1)$ and $K'$ $(\tau = -1)$ points can be written as $H^{\tau}_{\textrm{MG}}(\vec{k}) =v\vec{k}\cdot \sigma^{\tau} +U\mathbbm{1}_{2\times 2}$ with $\sigma^{\tau} = (\tau\sigma_x, \sigma_y)$ and  
\begin{align}
H^{\tau}_{\textrm{BG}}(\vec{k})= \left (
\begin{array}{cccc} 0 & v \pi^{\tau^*} & -v_4 \pi^{\tau^*} &  -v_3 \pi^{\tau}\\
             v \pi^{\tau} &\Delta & \gamma_1 & -v_4 \pi^{\tau*} \\
              -v_4 \pi^{\tau} & \gamma_1 
             & -U +\Delta & v \pi^{\tau*} \\
              -v_3 \pi^{\tau*} &  -v_4 \pi^{\tau}&  v \pi^{\tau}& -U  
             \end{array} \right),\label{eq:bilayerHam}
             \nonumber
\end{align}
with $\pi^{\tau} = \tau\vec{k}_x-i\vec{k}_y$. Here, $\sigma_i$, $i=x,y,z$, are the Pauli matrices in sublattice space. $T^{\tau}$ in Eq.~(\ref{eq:Ham}) describes the tunneling between the graphene monolayer and the upper layer of the graphene bilayer. In momentum space, it can be written as
$[T^{\tau}]_{\vec{k}',\vec{k}} = \sum_{j=1}^3 T^{\tau}_j\delta_{\vec{k}',\vec{k}-\vec{q}_j}$, with 
\begin{align}
T^{\tau}_j =\begin{pmatrix}
w_{AA} & w_{AB}e^{-i\tau(j-1)\frac{2\pi}{3}}  & 0 & 0 \\
w_{AB}e^{i\tau(j-1)\frac{2\pi}{3}} &  w_{AA} & 0 & 0 \\
  \end{pmatrix}.\nonumber 
\end{align}
Here, $T^{\tau}_j$ is modeled such that only the upper layer of the graphene bilayer, adjacent to the graphene monolayer in \figref{Fig0:Setup}, experiences the moir\'e potential. The interlayer tunneling processes are dominated by three momenta:  $\vec{q}_1 = K_{\theta}(0,-1), \ \vec{q}_2 = K_{\theta}(\frac{\sqrt{3}}{2},\frac{1}{2}), \ \vec{q}_3 = K_{\theta}(-\frac{\sqrt{3}}{2},\frac{1}{2})$, with $K_{\theta} = 8\pi \ \textrm{sin}(\theta/2)/{3a_0}$ and graphene lattice constant $a_0$~\cite{bistritzer2011moire,park2023network}. We have chosen the following values for the physical parameters in the above equations:  $v = 8.47 \times 10^5 \ \textrm{m/s}, \ v_3 = 9.19 \times 10^4 \ \textrm{m/s}, \ v_4 = 4.48 \times 10^4 \ \textrm{m/s} \ \gamma_1 = 361 \ \textrm{meV}, \ \Delta = 15 \ \textrm{meV}, \ w_{AA} = 80 \ \textrm{meV} \ \textrm{and} \ w_{AB} = 110 \ \textrm{meV}$. These were taken from DFT calculations previously reported in the literature~\cite{polshyn2020electrical, mccann2013electronic, park2020gate, park2023network, polshyn2022topological, rademaker2020topological,wang2024chern}. Unless stated otherwise, we have selected a twist angle $\theta = 1.21^{\circ}$ and an electrostatic potential $U = -40 \ \textrm{meV}$, which leads to two comparatively flat conduction minibands close to charge neutrality, with a finite indirect band gap separating them from the remote bands \cite{park2020gate}.
In this range of twist angles and displacement fields, the anomalous Hall effect as well as electrical switching of the magnetic order were observed in experiments on TMBG~\cite{polshyn2020electrical, chen2021electrically, polshyn2022topological, xu2021tunable}.  Although not indicated explicitly in \equref{eq:Ham}, all bands of the Hamiltonian in \equref{eq:Ham} are twofold spin degenerate. 

\begin{figure*}[t]
	\centering
	\includegraphics[width= 2.1\columnwidth]{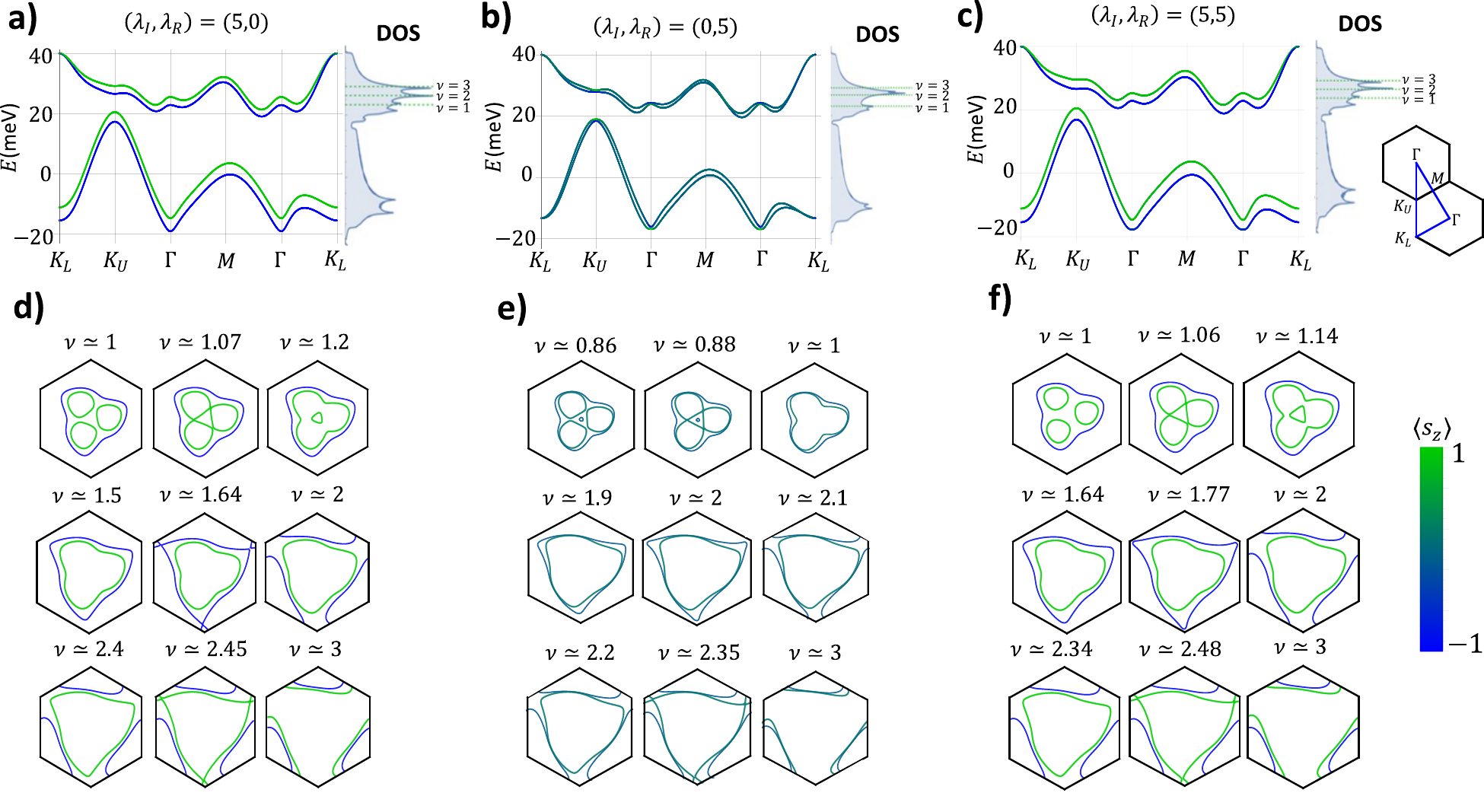}
	\caption{{\bf The single-particle miniband dispersions, DOS and Fermi surfaces for SOC-TMBG with different SOC parameters.}
    \justifying
    {(\textbf{a-c}) Single-particle band dispersion for the lowest conduction and valence minibands as well as the corresponding DOS. The dispersions and the DOS are for $K$ valley only. These data were obtained by diagonalizing Eq.\ (\ref{eq:Ham}) for $K$ valley. The different panels are for different SOC parameters: (\textbf{a}) $(\lambda_I, \lambda_R) = (5,0)$, (\textbf{b}) $(\lambda_I, \lambda_R) = (0,5)$, (\textbf{c}) $(\lambda_I, \lambda_R) = (5,5)$. The SOC parameters are in units of meV. The dispersions are plotted along the momentum path depicted in the inset of panel ($\textbf{c}$) covering the high symmetry points ($K_L\rightarrow K_U\rightarrow\Gamma\rightarrow M \rightarrow\Gamma\rightarrow K_L$). At each momentum, the miniband dispersion is color coded according to the spin expectation value $\langle s_z \rangle$. The color legend is shown on the right of panel ($\textbf{f}$). (\textbf{d-f}) Fermi contours for the  same SOC parameter sets as in panels (\textbf{a-c}) at integer fillings and near fillings where the Fermi surface topology changes due to the emergence of VHS. The color coding is again based on the spin expectation value $\langle s_z\rangle$ in the mBZ. 
	}}
	\label{Fig1:TMBGSOC}
\end{figure*}

This degeneracy is lifted once we take SOC into account induced by the TMD layer located on top of the TMBG. To this end, we add the following SOC Hamiltonian to the graphene monolayer  in Eq.\ (\ref{eq:Ham})
\begin{equation}
H_{\textrm{SOC}} =\lambda_I s_z\tau_z + \lambda_R (\sigma_x s_y\tau_z -\sigma_y s_x)\label{eq:SOCham},
\end{equation}
where $s_j, \tau_j$ are Pauli matrices in spin and valley space, respectively. $H_{\textrm{SOC}}$ is described by two parameters -- the strength of the Ising ($\lambda_I$) and Rashba ($\lambda_R$) SOC. The magnitude of the proximity-induced SOC from the TMD is mainly determined by the wavefunction overlap between the states of the TMD layer and those of the graphene monolayer~\cite{lin2022spin}. Previous DFT studies have shown that these SOC parameters are sensitive to several factors of the system, such as the type of TMD layer, the twist angle between the TMD layer and the graphene layer, the applied displacement field, and strain. The absolute values of $\lambda_I$ and $\lambda_R$ in the literature are reported to vary by as much as 2 meV ~\cite{fulop2021boosting, PhysRevB.100.085412, wang2016origin, li2019twist, gmitra2016trivial, gmitra2015graphene, sun2023determining}. For example, Ref.\ \cite{li2019twist} reported the SOC parameters as a function of twist angle between a graphene layer and a tungsten diselenide layer, from $0^{\circ}$ to $30^{\circ}$. The strength of the Ising and Rashba SOC were found to increase for twist angle up to $15^{\circ}\sim 20^{\circ}$ and then decrease for larger twist angle up to $30^{\circ}$. One experimental work reported much larger values of $\lambda_R$ on the order of $10$ meV, although this value strongly depends on the twist angle as well as the asymmetry of the dielectric environment \cite{sun2023determining}.

TMBG has lower symmetry than other graphene based moir\'e systems composed of layers with symmetric stacking order \cite{lin2022spin, tan2024topological, scammell2023tunable,PhysRevB.100.085109}. SOC-TMBG possesses spinful time-reversal symmetry as well as $C_{3z}$ rotational symmetry, but $C_{2z}$ symmetry is absent due to the geometry of the AB stacked graphene bilayer. Here,  $z$ refers to the axis perpendicular to the graphene plane. For heterostructures composed of twisted layers with symmetric stacking order and proximitized TMD layer, the twist angle can be tuned 
to preserve the $C_{2z}$ symmetry \cite{lin2022spin, tan2024topological, scammell2023tunable}, This is not possible in SOC-TMBG and, hence, a non-zero Ising SOC is allowed. Its presence reduces the SO(3) spin rotational symmetry down to SO(2). Rashba SOC breaks both the spin SO(3) and $C_{3z}$ symmetries, but it preserves the combination $C_{3z}^s = C_{3z} \otimes S_{\hat{z}2\pi/3}$, where $S_{\hat{n}\varphi}$ denotes a spin rotation by an angle $\varphi$ along $\hat{n}$. Finally, we note that the Hamiltonian for SOC-TMBG is invariant under a valley-$U(1)$ rotation, since it is fully diagonal in the valley index $\tau = \pm1$.

\subsection{Band structure of SOC-TMBG}
\label{spTMBG}
In this section, we discuss the dispersion of the lowest lying conduction and valence minibands,  whose energies are close to zero, the Fermi surfaces and the density of states (DOS) of SOC-TMBG. For a given band index $b$, flavor index $\alpha$   
and momentum $\vec{k}$, the band energy $\epsilon_{b\alpha}(\vec{k})$ is obtained by diagonalizing Eq.\ (\ref{eq:Ham}) with wavefunction $|u_{b\alpha}(\vec{k})\rangle$, which satisfies $H^{\tau}(\vec{k})|u_{b\alpha}(\vec{k})\rangle = \epsilon_{b\alpha}(\vec{k})|u_{b \alpha}(\vec{k})\rangle$. 
Note that when Rashba SOC is present, spin and band degrees of freedom are mixed. Therefore, the flavor $\alpha$ refers to both the spin and valley degrees of freedom when Rashba SOC is absent, and only to the valley degree of freedom when Rashba SOC is present.  

Figure \ref{Fig1:TMBGSOC}(a-c) display the dispersion of the lowest-lying conduction and valence minibands in the mBZ, as well as the DOS, for three different sets of the SOC parameters $(\lambda_I,\lambda_R)=(5,0),\ (0,5),\ (5,5)$ meV, respectively. 
In addition, Fig.\ \ref{Fig1:TMBGSOC}(d-f) display the Fermi surfaces for the $K$ valley (the $K'$ valley is related by time-reversal symmetry) for the same set of SOC parameters at different fillings, including integer fillings and fillings for which the Fermi energy is close to a van Hove singularity (VHS). The miniband dispersions as well as the Fermi surfaces at momentum $\vec{k}$ are color coded  according to the spin expectation value $\langle s_z \rangle$ and the color legend shown on the right of Fig.\ \ref{Fig1:TMBGSOC}(f). 

As anticipated, SOC causes a momentum-dependent spin splitting.  Note that at the $K_L$ symmetry point of the mBZ, the conduction miniband remains spin degenerate, irrespective of the SOC parameters, because these states belong to the lower graphene bilayer that does not experience SOC. In the remainder, we will refer to the combined SOC-split minibands lying above $E = 0$ in \figref{Fig1:TMBGSOC} and coming from the originally spin-degenerate lowest conduction minibands as the conduction bands. Analogously, the combined SOC-split highest minibands located below these conduction bands will be referred to as the valence bands. The splitting of the bands induced by Ising SOC is larger than that for Rashba SOC of the same magnitude.
To quantify this, we label the two split conduction bands as $c_i$, $i = 1,2$, and define the splitting $\Delta_{c_1c_2} $ as: 
\begin{align}
\Delta_{c_1c_2} = \underset{\vec{k}\in \textrm{mBZ}}{\textrm{Max}}|\epsilon_{c_1}(\vec{k})-\epsilon_{c_2}(\vec{k})|.\nonumber
\end{align}
For $(\lambda_I, \lambda_R) = (0,5)$, we obtain $\Delta_{c_1c_2}=1.68 \ \textrm{meV}$. This value is smaller than that for $(\lambda_I, \lambda_R) = (5,0)$ and $(\lambda_I, \lambda_R) = (5,5)$, where we obtain $\Delta_{c_1c_2} = 2.85 \ \textrm{meV}$  and $\Delta_{c_1c_2}= 3.08 \ \textrm{meV}$, respectively. 

Concerning the band topology \cite{fukui2005chern}, our choice of the electrostatic potential ($U = -40 \ \textrm{meV}$) and twist angle noted above lead for vanishing SOC to a conduction band with Chern number $2$ (for the $K$ valley), while the valence band has Chern number $-1$. For completeness, we further point out that reversing the sign of the displacement field leads to a conduction band with Chern number $-1$ and a valence band with Chern number $2$. For SOC values close to realistic DFT parameters, we find that the composite Chern number of the SOC-split conduction bands does not change \cite{fukui2005chern}, while the Berry-curvature distribution becomes sensitive to which type of SOC is present. The same applies to the valence bands. In particular, we find that the Berry curvature of both split conduction bands tends to be concentrated around the $\Gamma$ point in the mBZ when Rashba SOC becomes larger than Ising SOC. This can be explained by finite Rashba SOC-induced band mixing and the small band splitting especially at the $\Gamma$ point in mBZ. 

In addition, we observe that, depending on SOC, VHS can emerge close to integer fillings, therefore enhancing interaction effects. To visualize how the Fermi surface topology changes, we show the Fermi surface of the system around Van Hove fillings in \figref{Fig1:TMBGSOC}. For filling factor close to or smaller than $\nu = 1$, three pockets of Fermi energy contours from the lower conduction band start to touch each other, which leads to a change of Fermi surface topology. Additional changes of Fermi surface topology occur, when the triangular Fermi surface of the lower or upper conduction band touches the $K$ point of the mBZ at larger $\nu$.

\section{Hartree-Fock analysis for SOC-TMBG} \label{sec:HF}
Interaction effects play a crucial role for the low-energy physics when the electrons are filled to bands whose bandwidth is much smaller than the interaction strength. In our model, the single-particle bandwidth of TMBG is of the order of $20 \ \textrm{meV}$, whereas the Coulomb interaction strength is computed to be roughly of the order of 100 meV (see below). Based on this estimation, we expect complex, symmetry-broken phases to emerge in TMBG; indeed, there are already multiple theoretical studies that report possible correlated states in TMBG for different parameter regimes \cite{wang2024chern, rademaker2020topological,wilhelm2023non, polshyn2022topological, song2025fractional}, although almost all of these works focus on integer fillings with only a few exceptions \cite{wilhelm2023non, polshyn2022topological, song2025fractional}. In this section, we study the possible correlated states of SOC-TMBG for all positive integer and half-integer fillings and include the effect of SOC, by performing self-consistent Hartree-Fock (HF) calculations. We allow our HF solutions to also capture the translation symmetry broken states, which effectively enlarges the unit cell. In particular, motivated by the experimental result \cite{polshyn2022topological} and the exact diagonalization calculations \cite{wilhelm2023non, song2025fractional} (without SOC), we only consider momentum transfer vectors corresponding to the three $M$ points in the mBZ. 

\subsection{Interacting Hamiltonian and HF}
To outline the basic methodology, let us first define the band creation operator $c^{\dagger}_{\vec{k}b\alpha}$, which creates an electron at momentum $\vec{k}$, with band $b$, and of flavor $\alpha$, with corresponding single-particle energy $\epsilon_{b\alpha}(\vec{k})$. By construction, the single-particle Hamiltonian $H_S$ is already diagonal in this basis and can be rewritten as
\begin{align}
H_{S} = \sum_{\vec{k}\in \textrm{mBZ}}\sum_{b,\alpha}\epsilon_{b\alpha}(\vec{k})c^{\dagger}_{\vec{k}b\alpha}c^\pdagger_{\vec{k} b\alpha}  \label{eq:BM}.
\end{align}
The density-density Coulomb interaction reads
\begin{align}
\hspace{-0.3cm}H_V=\frac{1}{2A}\sum_{\vec{q}}V(\vec{q})\rho(-\vec{q})\rho(\vec{q})
\label{eq:CouHam},
\end{align}
with total system size $A$. Here we define the density operator $\rho(\vec{q})$ and the form factor $\Lambda_{\vec{q}}(\vec{k})$ as $\rho(\vec{q})=\sum_{\vec{k}\in \textrm{mBZ}}c^{\dagger}_{\vec{k}}\Lambda_{\vec{q}}(\vec{k})c^\pdagger_{\vec{k}+\vec{q}} $ and $[\Lambda_{\vec{q}}(\vec{k})]_{b_1b_2\alpha\beta} =  \langle{u_{b_1\alpha}({\vec{k}})}|u_{b_2\beta}(\vec{k}+\vec{q})\rangle$ for given band indices $b_1, b_2$ and flavor indices $\alpha, \beta$, respectively. In addition, we use the double-gate screened Coulomb potential $V(\vec{q}) = \frac{e^2}{2\epsilon \epsilon_0 |\vec{q}|}\textrm{tanh}(|\vec{q}| d)$, with screening length $d$ and dielectric constant $\epsilon$. We have checked that using a single-gate screened Coulomb potential ($V(\vec{q}) = \frac{e^2}{2\epsilon \epsilon_0 |\vec{q}|}(1-e^{-2|\vec{q}|d})$), instead of a double gate interaction, does not qualitatively change the main result of our work. For the interaction parameters, we set $d = 20 \ \textrm{nm}$ and $\epsilon = 10$. 

For computational efficiency, we project the interacting Hamiltonian $H_{\textrm{int}}=H_{S}+H_V$ onto the small set of bands (active bands) whose energies are close to zero. We consider the cases with Fermi level inside the conduction bands because they are very narrow and isolated for our choice of parameters and also this is where signatures of correlated states were observed. In the main text, we project  $H_{\textrm{int}}$ only onto these conduction bands. In \appref{WithValenceBands}, we then, as a second step, generalize by including both the conduction and the valence bands. We note that in the first projection scheme, charge neutrality ($\nu=0$) corresponds to an entirely empty active band, while for the latter, it corresponds to fully filled valence bands. 

As usual in HF, we treat the four-fermion Coulomb interaction $H_{V}$ by performing all possible particle-number-conserving mean-field decouplings. It then becomes an effectively quadratic Hamiltonian,
\begin{align}
H_{\textrm{C}}= \sum_{\vec{k}\in \textrm{mBZ}, \vec{Q}_i\in \mathcal{Q}} c^{\dagger}_{\vec{k}}h_{\textrm{HF}}[P(\vec{k},\vec{Q}_i)]c^\pdagger_{\vec{k}-\vec{Q}_i} \label{eq:Coulomb},
\end{align}
with Hamiltonian function $h_{\textrm{HF}}$ that depends on the correlator matrix $P(\vec{k},\vec{Q}_i)=\langle c^{\dagger}_{\vec{k}}c_{\vec{k}+\vec{Q}_i}\rangle$, see \appref{AppendixHF}. Physically, $\vec{Q}_i$ corresponds to the momentum transfer vector describing potential moir\'e translation symmetry breaking; we further introduce $\mathcal{Q}$ as the set of possible momentum vectors $\vec{Q}_i$ that we allow in our HF calculation to develop. 
For later reference, we also define the combined correlator matrix $[\vec{P}(\vec{k}, \mathcal{Q})]_{ij} = P(\vec{k}+\vec{Q}_i, -\vec{Q}_j)$ for given momentum indices $i,j$. 

\subsection{Subtraction scheme}
The above-mentioned model parameters we use, such as Dirac velocity and tunneling strength, are taken from the references \cite{park2020gate, polshyn2020electrical, polshyn2022topological,wang2024chern, rademaker2020topological, park2023network, mccann2013electronic}; they are obtained from DFT calculations and thus already contain some interaction effects. To prevent the double counting of interactions, we subtract a reference density matrix $\vec{P}_0(\vec{k}, \mathcal{Q})$ from the correlator $\vec{P}(\vec{k}, \mathcal{Q})$, which incorporates the interaction effects already included in DFT parameters. In our work, we choose an 'average' scheme, where we set $\vec{P}_0(\vec{k}, \mathcal{Q}) = \mathbbm{1}/2$ for active bands, $\vec{P}_0(\vec{k}, \mathcal{Q}) = \mathbbm{1}$ for occupied remote bands and $\vec{P}_0(\vec{k}, \mathcal{Q}) = 0$ for unoccupied remote bands. We note that our results are expected to be qualitatively unaffected when other common reference schemes are used \cite{parker2021field}. 

Including both contributions, $H_S$ and $H_{\textrm{C}}$ from Eqs.~(\ref{eq:BM}) and (\ref{eq:Coulomb}), respectively, the full HF Hamiltonian can be written as
\begin{align}
\hspace{-0cm}H_{\textrm{HF}} =\sum\limits_{\vec{k}\in \textrm{mBZ}_{\mathcal{Q}}} \textbf{c}^{\dagger}_{\vec{k}\mathcal{Q}}\textbf{h}_{\textrm{HF}}[\vec{k}, \mathcal{Q}, \vec{P}_{1}(\vec{k}, \mathcal{Q})]\textbf{c}^\pdagger_{\vec{k}\mathcal{Q}}\label{eq:HFHam},
\end{align}
 with the new spinor $\textbf{c}_{\vec{k}\mathcal{Q}} \equiv (c_{\vec{k}} \  \cdots \ c_{\vec{k}+\vec{Q}_i} \ \cdots)^T$ and HF Hamiltonian functional $\textbf{h}_{\textrm{HF}}$. The momentum summation of Eq. (\ref{eq:HFHam}) is performed over the folded moir\'e Brillouin zone ($\textrm{mBZ}_{\mathcal{Q}}$), which depends on the choice of $\mathcal{Q}$ (see also sec.\  \ref{choicesec}). In addition, for $R = 1, 2$, we define $\vec{P}_R(\vec{k},\mathcal{Q})\equiv \frac{1}{R}\vec{P}(\vec{k},\mathcal{Q})-\vec{P}_0(\vec{k},\mathcal{Q})$. From \equref{eq:HFHam}, the HF energy $E_{\textrm{HF}}^{\mathcal{Q}}$ for given correlator $\vec{P}(\vec{k}, \mathcal{Q})$ can then be compactly written as  
\begin{align}
\hspace{0cm}E^{\mathcal{Q}}_{\textrm{HF}} = \sum_{\vec{k}\in\textrm{mBZ}_{\mathcal{Q}}}\textrm{Tr}[\vec{P}(\vec{k},\mathcal{Q})\textbf{h}^T_{\textrm{HF}}[\vec{k}, \mathcal{Q},\vec{P}_2(\vec{k}, \mathcal{Q})]] \label{eq:HFener}.
\end{align}
We start with several initial random ans\"atze of $\vec{P}(\vec{k}, \mathcal{Q})$, repeat the iterative calculation until convergence is reached, and pick the solution which gives the minimum value of $E^{\mathcal{Q}}_{\textrm{HF}}$. See \appref{AppendixHF} for a detailed derivation of Eqs.~(\ref{eq:Coulomb}), (\ref{eq:HFHam}) and (\ref{eq:HFener}) with different $\mathcal{Q}$.

\begin{figure*}[t]
	\centering
	\includegraphics[width= 2.1\columnwidth]{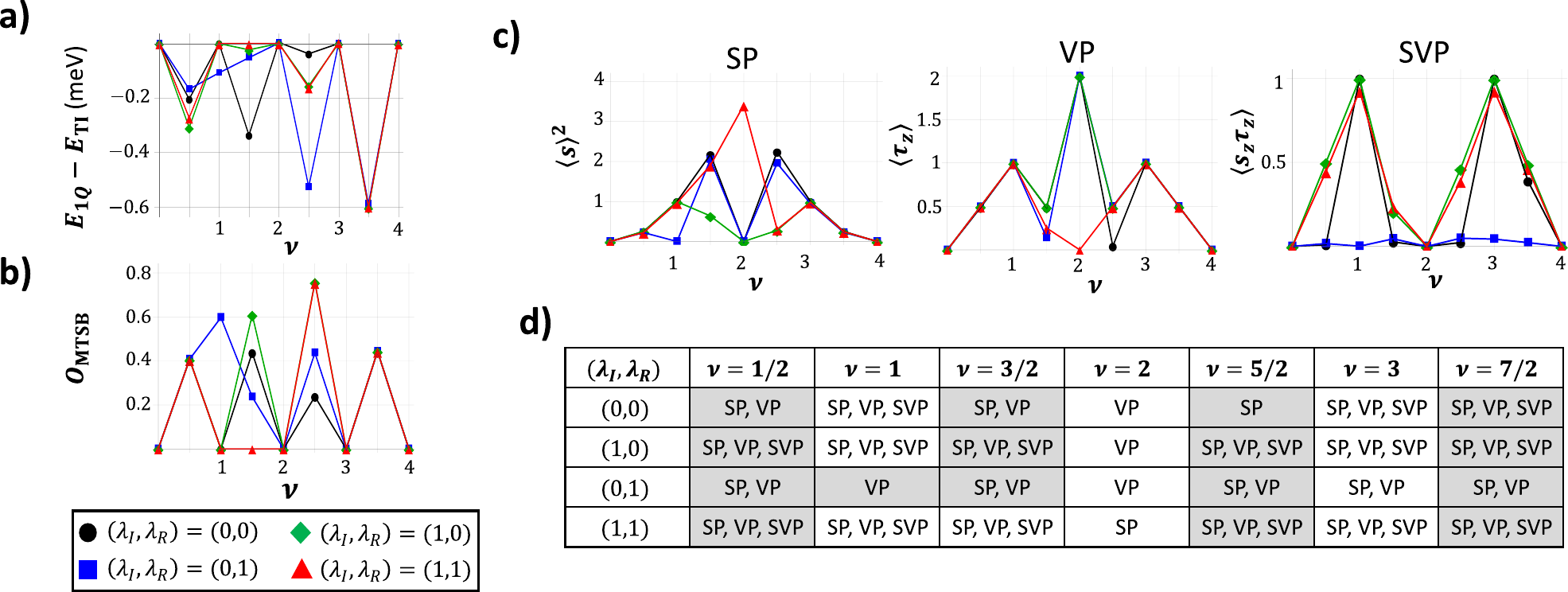}
	\caption{{\bf HF data for translation-invariant (TI) and $1Q$ states} 
\justifying
    with different SOC parameters. (\textbf{a}) HF energy difference between $1Q$ and TI states per moir\'e unit cell as a function of filling-factor $\nu$ with different SOC parameters.
    (\textbf{b}) Filling-factor dependence of moir\'e translation symmetry breaking order parameter $O_{\textrm{MTSB}}$ with different SOC parameters.
     (\textbf{c}) Filling-factor dependence of $\langle\vec{s}\rangle^2=\langle s_x \rangle^2+\langle s_y \rangle^2+\langle s_z \rangle^2$, $\langle \tau_z \rangle$ and $\langle s_z \tau_z \rangle$ with different SOC parameters. See Eq.\ (\ref{eq:SDWorder}) and Eq.\  (\ref{eq:expectation}) for the definitions of order parameters. (\textbf{d}) Table showing which flavor polarized states emerge for each HF ground state at different fillings $\nu$ and SOC parameters. The unit of SOC parameters is meV. Here, VP, SP, and SVP stand for valley polarized, spin polarized, and spin-valley polarized order, respectively. The gray (non)-shaded sectors of the table denote $1Q$ (TI) states. The SOC parameters are in units of meV. The system size is $12\times12$.
	}
	\label{Fig4:HFTMBG}
\end{figure*}

\subsection{Choice of wave vector and SOC parameters}\label{choicesec}
Motivated by the experimental signatures of translation symmetry breaking at $\nu = 7/2$ and previous theoretical work \cite{wilhelm2023non, polshyn2022topological, dong2023many, song2025fractional}, we perform HF calculations for the three possible cases 
\begin{enumerate}
    \item $\mathcal{Q} =\{\Gamma\}$
    \item $\mathcal{Q} =\{\Gamma, \  M_1\}$ 
    \item $\mathcal{Q} =\{\Gamma, \ M_1, \  M_2, \  M_3\}$,
\end{enumerate}
where the variational space is being extended with respect to the momentum vectors that are included. There are three inequivalent $M$ points in the mBZ. See the inset of Fig.~\ref{Fig1:TMBGSOC} for the location of one of them; the other two can be obtained from it by a $\pm120^\circ$ rotation. The first case can only capture translation invariant (TI) solutions. Meanwhile, the second and third cases also contain translation-symmetry-broken solutions with, respectively, at most one pitch vector given by $M_1$ (referred to as ``$1Q$ states'' in the following) or up to a superposition of all three $M$ points, $M_1, M_2, M_3$. When a superposition of all three of them is found, we refer to such a state as a ``$3Q$ state''. Note that, as a result of the variational principle, the optimal $E^{\mathcal{Q}}_{\textrm{HF}}$ cannot increase when additional vectors are included in $\mathcal{Q}$. 

In addition, we define mBZ$_{1Q} \ (\mathcal{Q}=\{\Gamma, M_1\})$, mBZ$_{3Q} \ (\mathcal{Q}=\{\Gamma, M_1, M_2, M_3\}) $ as the folded Brillouin zone corresponding to $1Q$ and $3Q$ states, respectively. More explicitly, mBZ$_{1Q}$ (mBZ$_{3Q}$) is the two-fold (four-fold) reduced mBZ associated with the momentum transfer given by single (all three) $M$ point(s). See \appref{FoldedBZ} for details of the numerical implementation of the mBZ$_{\mathcal{Q}}$ grids.

For the various HF solutions, we focus on the state with the lowest energy for each integer and half-integer fillings in the range $ 0 \leq \nu \leq 4$. In addition, our discussions will focus on how the SOC parameters affect the correlated physics of SOC-TMBG, especially regarding the spin configurations. For the SOC parameters, we consider four possible combinations 
$$(\lambda_I, \lambda_R) = (0,0), \ \ (1,0), \ \ (0,1), \ \ (1,1) \ \textrm{meV}$$\,.
They already have a strong impact on the correlated ground states, as we will see in the following.

\section{HF analysis in conduction bands: $1Q$ state}
\label{Sec:HFconductionBands}
 As anticipated above, we begin our HF analysis by only keeping the in total four conduction bands (including the valley index) that cross the Fermi level. For the chosen model parameters, the single-particle band structure of SOC-TMBG shows that there is not only a finite indirect band gap between the conduction and valence bands, but they are also separated energetically from the upper and lower remote bands. As such, our projection onto the conduction band is well defined and will allow us to first develop a clearer physical picture. As we show in \appref{WithValenceBands}, including additional bands in the HF analysis will not lead to any changes in the symmetries of the stabilized ground states for $\nu>0$.
 
\subsection{Energy comparison between TI and ${1Q}$ states}\label{EnCompTI1Q}
We first focus on TI and $1Q$ ans\"atze, by setting $\mathcal{Q} = \{\Gamma\}$ and $\mathcal{Q} = \{\Gamma, \   M_1\}$, respectively, and study all possible integer and half-integer fillings between $\nu=0$ (completely empty system) and $\nu=4$ (all active bands are filled). To distinguish between TI and $1Q$ states qualitatively, we quantify the degree of translational symmetry breaking on the moiré scale by defining
\begin{align}
O_{\textrm{MTSB}} =\frac{1}{N_{1Q}}\sum_{\vec{k}\in \textrm{mBZ}_{1Q}}\sqrt{\textrm{Tr}[P(\vec{k},M_1)P^{\dagger}(\vec{k},M_1)]} \label{eq:SDWorder},
\end{align} 
where $N_{1Q}$ is the number of points of the $\textrm{mBZ}_{1Q}$ grid. This quantity extracts the weight of the off-diagonal components of the entire correlator matrix $P(\vec k,\mathcal Q=\{\Gamma,M_1\})$ with respect to the electronic momenta $\vec{k}$ and $\vec{k} + \vec{M}_1$. Thus, it captures spatial modulations on the moiré scale. We distinguish the TI and $1Q$ state by checking whether the state has zero or nonzero $O_{\textrm{MTSB}}$.

Figure~\ref{Fig4:HFTMBG}(a) shows the filling-factor dependence of the energy difference per moir\'e unit cell between the two self-consistent HF solutions obtained from the TI and $1Q$ ans\"atze for different SOC parameters. In addition, \figref{Fig4:HFTMBG}(b) shows the filling-factor dependence of $O_{\textrm{MTSB}}$ for these ground states. As anticipated, we first see that without SOC, both TI and $1Q$ ans\"atze converge to  TI states at integer fillings with zero $O_{\textrm{MTSB}}$. This feature is qualitatively robust even in the presence of SOC, except at $\nu=1$ and $(\lambda_I, \lambda_R) = (0,1) \ \textrm{meV}$, where a $1Q$ state has lower energy per moir\'e unit cell than the TI state (by approximately 0.11 meV). In contrast, for half-integer fillings, a  $1Q$ state with nonzero $O_{\textrm{MTSB}}$ becomes energetically favored over TI states. We also see that the presence of SOC quantitatively affects the energy difference between TI and $1Q$ states. Finally, we note that the HF ground states for all considered fillings and SOC parameters are insulating phases with a finite band gap.

\begin{figure*}[t]
	\centering
	\includegraphics[width= 1.6\columnwidth, height= 1\columnwidth]{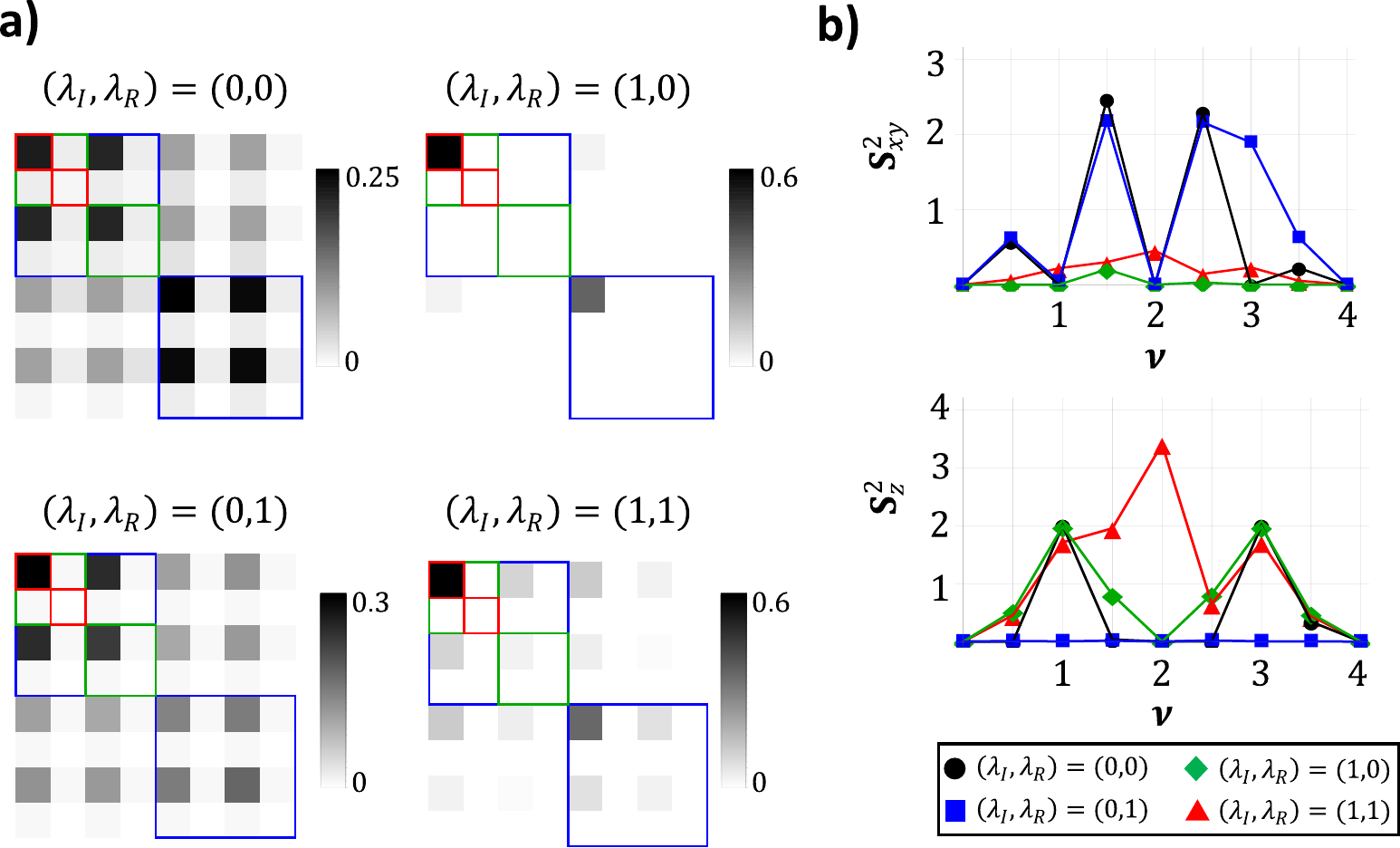}
	\caption{{\bf Correlator matrix and spin order parameter of $1Q$ states} (\textbf{a}) 
    \justifying
    Correlator matrix structure with an absolute value $|\tilde{\textbf{P}}_{1Q}|$ at filling factor $\nu = 1/2$ with different SOC parameters. The matrix structure of $\tilde{\textbf{P}}_{1Q}$ is built such that the smallest red-outlined block denotes valley flavor, the next green-outlined block denotes spin (or band depending on presence of Rashba SOC) flavor, and the largest blue-outlined block denotes the additional degrees of freedom from broken moir\'e translation symmetry.  (\textbf{b}) Filling-factor dependence of in-plane $\textbf{S}^2_{xy}$ and out-of-plane $\textbf{S}^2_z$ spin order parameter for the HF ground state with different SOC parameters. The SOC parameters are in units of meV. The system size is $N=12\times12$.
	}
	\label{Fig5:HFTMBG2}
\end{figure*}

\subsection{Spin and/or valley polarized states}\label{SVPSs}
To study the nature of the correlated states, we focus on the converged correlator matrices $\vec{P}_{\textrm{TI}}(\vec{k}) \equiv\vec{P}(\vec{k}, \mathcal{Q}=\{\Gamma\})$ and $\vec{P}_{1Q}(\vec{k}) \equiv \vec{P}(\vec{k}, \ \mathcal{Q} = \{\Gamma,  M_1\})$ of TI and $1Q$ states, respectively, obtained from our self-consistent HF calculation. We decompose $\vec{P}_{\textrm{TI}}(\vec{k})$ into valley and spin degrees of freedom, and $\vec{P}_{1Q}(\vec{k})$ into valley, spin, and additional degrees of freedom from broken moir\'e translation symmetry. We denote the associated Pauli matrices by $\tau_i$, $s_i$, and $o_i$, respectively. In particular, for each TI and $1Q$ states, we sum the correlator over the momentum $\vec{k}\in\textrm{mBZ}$ or $\vec{k}\in\textrm{mBZ}_{1Q}$  and define its average as
\begin{align}
\tilde{\textbf{P}}_{\mu} \equiv \frac{1}{N_{\mu}}\sum_{\vec{k}\in \textrm{mBZ}_{\mu}}\vec{P}_{\mu}(\vec{k})\label{correlator}
\end{align}
with $\mu= \textrm{TI}, \ 1Q$ and $N_{\mu}$ as the number of momentum points on the mBZ$_{\mu}$ grid. Here we use $\textrm{mBZ}_{\text{TI}}$ to denote the original, non-folded mBZ. Equation~(\ref{correlator}) allows us to compactly write the expectation values of arbitrary combinations of Pauli matrices $O$ as 
\begin{align}
 \langle O\rangle _{\mu} = \textrm{Tr}[\tilde{\textbf{P}}_{\mu}O]. \label{eq:expectation}
\end{align}
For each integer and half-integer filling, we pick the HF ground state $\textrm{GS}\in \{\textrm{TI}, 1Q\}$ and define $\langle O\rangle = \langle O\rangle_{\textrm{GS}}$.

Using Eq.\ (\ref{eq:expectation}), we compute the following expectation values: $ \langle\vec{s}\rangle^2 \equiv \langle s_x \rangle^2 + \langle s_y\rangle^2+\langle s_z \rangle^2$, $\langle \tau_z \rangle$, $\langle s_z \tau_z \rangle$ for each positive integer and half-integer filling-factor, which shows whether the correlated states correspond to spin polarized order (SP), valley polarized order (VP), spin-valley polarized order (SVP), or non-polarized order (NP). More explicitly, we use the following criteria to define SP, VP, SVP, and NP as
\begin{enumerate}
    \item SP: $\langle\vec{s}\rangle^2 \neq 0$
    \item VP: $\langle \tau_z\rangle \neq 0$ 
    \item SVP: $\langle s_z \tau_z \rangle \neq 0$
    \item NP: $\langle\vec{s}\rangle^2 = \langle \tau_z\rangle =\langle s_z \tau_z \rangle  = 0$
\end{enumerate}
Note that we only consider out-of-plane SVP, which has the same form as the spin-valley locking term of the Ising SOC in Eq.~\eqref{eq:SOCham}. This implies that if Ising SOC is non-zero, $\langle s_z \tau_z \rangle$ is no longer a symmetry-breaking order parameter, and any additional interaction-induced SVP can only manifest as a crossover rather than a true phase transition.

Fig.\ \ref{Fig4:HFTMBG}(c) shows the filling-factor dependence of $ \langle\vec{s}\rangle^2$, $\langle \tau_z\rangle$ and $\langle s_z\tau_z\rangle$ for different values of SOC parameters. Based on Fig.\ \ref{Fig4:HFTMBG}(c) and the criteria mentioned above, the table in Fig.\ \ref{Fig4:HFTMBG}(d) shows which flavor degrees of freedom are polarized or not for different fillings $\nu$ and SOC parameters. We see that Ising SOC induces SVP, while Rashba SOC suppresses SVP. This shows that these SOC terms have rather different consequences for the correlated physics.

The competition between Ising and Rashba SOC to induce or suppress SVP in the correlated states can be understood intuitively from their form in Eq.\ (\ref{eq:SOCham}). The Ising term is proportional to the SVP parameter $s_z\tau_z$, thus couples linearly to it, and hence strongly favors SVP order in the ground state. 
However, Rashba SOC reduces the SVP of the correlated states, since the ground-state expectation values of terms involving $s_x$ and $s_y$ must be non-zero in order to couple to $\lambda_R$ in linear order.

\subsection{Spin configuration}
Let us next take a closer look at how the SOC affects the spin configuration of the correlated states obtained within our HF analysis. In Fig.\ \ref{Fig5:HFTMBG2}, we visualize the matrix structure of $\tilde{\textbf{P}}_{1Q}$ defined in Eq.\ (\ref{correlator}) for fixed filling $\nu = 1/2$ and different SOC parameters. At this filling, a $1Q$ state has lower energy than a TI state. First, we clearly see that Ising SOC only allows the spin-diagonal matrix element to be nonzero, which corresponds to out-of-plane spin components (we use the $s_z$ basis). This can again be understood through the spin-valley locking term $\sim \lambda_I s_z\tau_z$ from the Ising SOC Hamiltonian. For a more quantitative statement, we define the magnitudes $\mathbf{S}^2_{\alpha}$ of the different spin components ($\alpha = x,y,z$) for TI and $1Q$ states, respectively, as
\begin{align}
(\mathbf{S}_{\alpha}^2)_{\textrm{TI}} = \sum\limits_{j}\langle s_{\alpha}\tau_j \rangle ^2_{\textrm{TI}}, \ \   (\mathbf{S}_{\alpha}^2)_{1Q} = \frac{1}{2} \sum\limits_{j,k}\langle s_{\alpha}\tau_jo_k \rangle ^2_{1Q}. \label{eq:spin}
\end{align}
Note the additional factor of one-half for the $1Q$ spin order parameter, which accounts for the two additional degrees of freedom arising from broken moir\'e translation symmetry associated with a single $M$ point. In addition, we used the short-cut $\sum_{i,j,k,\cdots} \equiv\sum_{i,j,k,\cdots \in\{0,x,y,z\}}$. 

Fig \ref{Fig5:HFTMBG2}(b) shows the filling-factor dependence of the in-plane component of the spin order parameter $\textbf{S}^2_{xy} = (\textbf{S}^2_{xy})_{\textrm{GS}}\equiv(\textbf{S}^2_x)_{\textrm{GS}}+(\textbf{S}^2_y)_{\textrm{GS}}$ and of its out-of-plane analogue $\textbf{S}^2_z = (\textbf{S}^2_z)_{\textrm{GS}}$ for HF ground state $\textrm{GS}\in\{\textrm{TI}, 1Q\}$ with different SOC parameters. Our data clearly show that Ising SOC suppresses the in-plane spin components, due to the spin-valley locking term $(\sim \lambda_I s_z\tau_z)$, while Rashba SOC suppresses the out-of-plane spin configuration, due to the $s_x, s_y$ term in  the Rashba SOC Hamiltonian.

\section{Ginzburg-Landau analysis}\label{GLSection}
Before we go through the HF data for the $3Q$ ansatz, we study the expected $3Q$ states of SOC-TMBG phenomenologically by performing a Ginzburg-Landau (GL) analysis from the symmetry point of view, focusing on the spin order. More specifically, this section serves two purposes: first, it introduces nomenclature for the different possible states and their symmetries in a simplified setting; and second, it provides intuition about the fate of the TAF \cite{wilhelm2023non} when SOC is included.

\begin{figure*}[t]
	\centering
	\includegraphics[width= 2.1\columnwidth]{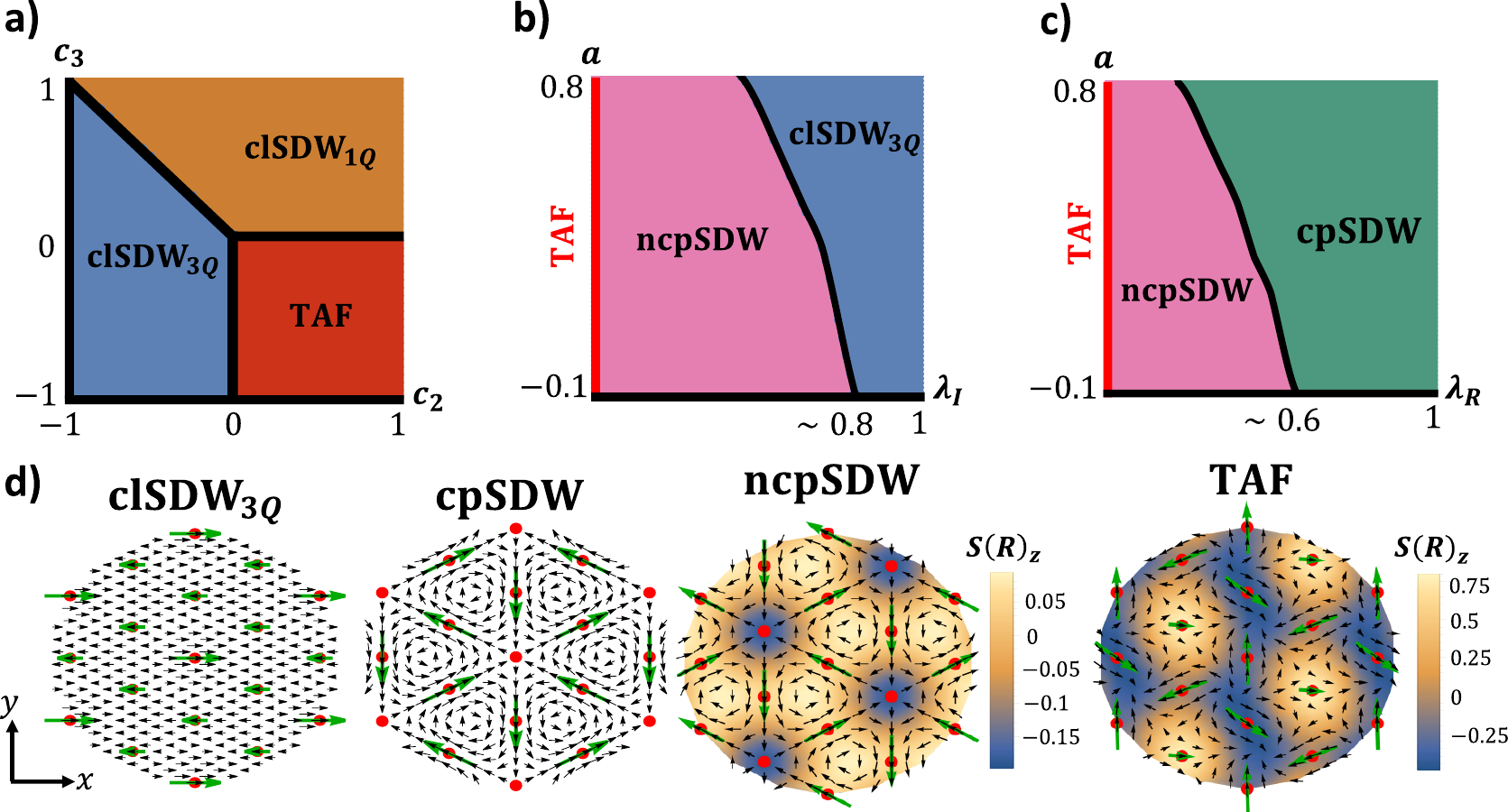}
	\caption{{\bf GL phase diagram and the ground state configuration}
    \justifying
    (\textbf{a}) GL phase diagram for zero SOC for different parameter sets $(c_2, c_3)$. (\textbf{b}) GL phase diagram for nonzero Ising SOC for different parameter sets $(\lambda_I, a)$. (\textbf{c}) GL phase diagram for nonzero Rashba SOC for different parameter sets $(\lambda_R, a)$. (\textbf{d}) Configuration of net spin expectation value $\vec{S}(\vec{R})$ for a collinear SDW ($\textrm{clSDW}_{3Q}$) state, coplanar SDW (cpSDW) state, non-coplanar SDW (ncpSDW) state, and tetrahedral antiferromagnetic (TAF) state in moir\'e real space, which are obtained from a GL analysis. For the figures, we choose (i) a $\textrm{clSDW}_{3Q}$ state with $\vec{\phi}_1 = \vec{\phi}_2=\vec{\phi}_3\neq0$, (ii) a cpSDW state with $m_{12}=m_{23}=m_{31}=-\frac{1}{2}$ and (iii) a ncpSDW state with $m_{12}=-m_{23}=-m_{31}=-\frac{1}{2}, \ \chi_{\textrm{GL}}\simeq0.38$. Black arrows denote $(\vec{S}(\vec{R})_x, 
    \vec{S}(\vec{R})_y)$ at position $\vec{R}$, red dots denote AA sites in moir\'e real space and green arrows denote the direction of $(\vec{S}(\vec{R}=\vec{R}_{AA})_x, 
     \vec{S}(\vec{R}=\vec{R}_{AA})_y)$ at AA sites. For ncpSDW and TAF, we also show the $z$ component of net spin expectation value $ \vec{S}(\vec{R})_z$ for each position $\vec{R}$, which 
    illustrates the non-coplanar nature of the spin configurations.}
	\label{Fig14:GLanalysis}
    \end{figure*}

\subsection{Symmetry analysis}\label{symanl}
To start with, we define the three-component real-valued order parameter $\vec{\phi}_j = (\phi^x_j, \phi^y_j, \phi^z_j)^T$ with $j=1,2,3$ as the index of the three $M_j$ points in the mBZ, and $x, y, z$ as the components in spin space. The spin expectation value in moiré unit cell $\vec{R}$ is then given by
\begin{equation}
  \vec{S}(\vec{R}) = \sum_{j=1}^3 \vec{\phi}_j \cos (\vec{M}_j\cdot\vec{R}). \label{SpinExpGL}
\end{equation}
For simplicity, we choose $\vec R=0$ to correspond to an AA site of the moir\'e lattice. To classify the different states using simple quantities, we define the scalar products $m_{ij}$ encoding the angle between two order parameters and the normalized scalar spin chirality $\chi_{\textrm{GL}}$ as a measure for the non-coplanarity of the spin texture as
\begin{equation}
     m_{ij}= \frac{\vec{\phi}_i \cdot \vec{\phi}_j}{|\vec{\phi}_i||\vec{\phi}_j|}, \,\,\, \chi_{\textrm{GL}} = \frac{|\vec{\phi}_1\cdot(\vec{\phi}_2 \times \vec{\phi}_3)|}{|\vec{\phi}_1||\vec{\phi}_2||\vec{\phi}_3|}. \label{mAndChiDefinition}
\end{equation}
This allows us to classify the following five states, which we refer to as collinear SDW  (clSDW$_{1Q}$, clSDW$_{3Q}$), non-coplanar SDW (ncpSDW), coplanar SDW (cpSDW), and tetrahedral antiferromagnet (TAF), defined as follows
\begin{enumerate}
    \item $\textrm{clSDW}_{1Q}: \vec{\phi}_1\neq 0, \ \vec{\phi}_2 = \vec{\phi}_3 = 0$, $m_{ij}= \chi_{\rm GL}=0$
    \item $\textrm{clSDW}_{3Q}: \vec{\phi}_1 = \pm\vec{\phi}_2 = \vec{\phi}_3 \neq 0$, $|m_{ij}|=1, \chi_{\rm GL}=0$
    \item cpSDW : $m_{12}=-\frac{1}{2}, m_{23}= m_{31} \pm \frac{1}{2}, \ \chi_{\textrm{GL}} = 0$
     \item $\textrm{ncpSDW}: m_{ij} \neq 0, \ 0<\chi_{\textrm{GL}} < 1,$
\item $\textrm{TAF}: |\vec{\phi}_1|  =  |\vec{\phi}_2| = |\vec{\phi}_3| \neq 0, \ \chi_{\textrm{GL}} = 1.$\end{enumerate}
\noindent See Fig.~\ref{Fig14:GLanalysis}(d) for a schematic depiction. Three comments are in order. First, among those states, cpSDW forms a vortex-antivortex lattice, while ncpSDW and TAF correspond to a skyrmion lattice; being coplanar, the first state is characterized by a vanishing skyrmion density $\mathbf{U}(\vec{R}) =|\vec{S}(\vec{R})\cdot (\partial_x \vec{S}(\vec{R})\times \partial_y \vec{S}(\vec{R}))| = 0$, while the latter two states have nonzero net skyrmion density $\mathbf{U} \equiv\int_{\vec{R}\in \textrm{UC}} d^2\vec R \,\mathbf{U}(\vec{R}) \neq 0$ with the integral going over a moiré unit cell. 
More specifically, for ncpSDW and TAF states, $\mathbf{U}(\vec{R})$ becomes finite at every position $\vec{R}$ except at AA sites (denoted as $\vec{R}_{AA}$), which is the center of a double vortex with winding number 2 of 
the $x,y$ component of $\vec{S}(\vec{R}=\vec{R}_{AA})$ ($\partial_x\vec{S}(\vec{R} = \vec{R}_{AA})=\partial_y\vec{S}(\vec{R} = \vec{R}_{AA})=0$) and therefore leads to zero skyrmion density ($\mathbf{U}(\vec{R}=\vec{R}_{AA})=0$). We further note that a spin-density modulation, as described by clSDW$_{1Q}$ and clSDW$_{3Q}$ states, automatically induces a charge modulation in the presence of our SOC terms, as confirmed in our HF analysis (see \secref{chiprod} below). Third, we note that the two different signs in the above list for $\textrm{clSDW}_{3Q}$ and cpSDW states simply correspond to symmetry-related spin configurations and thus constitute representatives of the same state. In detail, the configurations of two $\textrm{clSDW}_{3Q}$ states are related by a moir\'e-scale translation, whereas the two types of cpSDW states are related by a moir\'e-scale  translation combined with a spin rotation.

The symmetries present in our system for zero SOC are (i) $120^{\circ}$ rotation symmetry $C_{3z}$, (ii) translation symmetry $T_j$, (iii) time reversal symmetry $\theta$, and (iv) SO(3) spin rotational symmetry with $S_{\hat{n}\varphi}$ denoting a rotation by angle $\varphi$ along $\hat{n}$ in spin space. The order parameter $\vec{\phi}_j$ transforms for each symmetry as
\begin{enumerate}
    \item  $C_{3z}$: $\vec{\phi}_j \rightarrow \vec{\phi}_{j+1} \ (\vec{\phi}_4 \equiv \vec{\phi}_1)$
    \item $T_j$: $\vec{\phi}_{j'} \rightarrow (-1)^{\delta_{jj'}} \vec{\phi}_{j'}$ 
    \item $\theta$: $ \vec{\phi}_j \rightarrow -\vec{\phi}_j$
    \item $S_{\hat{n}\varphi}$ : $ \vec{\phi}_j \rightarrow R_{\hat{n}\varphi}\vec{\phi}_j$,
\end{enumerate}
where $R_{\hat{n}\varphi} \in \textrm{SO(3)}$ is the matrix representation of a rotation by angle $\varphi$ along $\hat{n}$.

When SOC is included, the Ising SOC reduces the SO(3) symmetry group to SO(2). In addition, Rashba SOC breaks $C_{3z}$ and $S_{\hat{n}\varphi}$ as separate symmetries, but preserves the combined $C_{3z}^s = C_{3z} \otimes S_{\hat{z}2\pi/3}$ symmetry, due to the hybridization term between spin and orbital degrees of freedom. 

\subsection{GL free-energy}
With these constraints at hand, we write the symmetry-allowed GL free-energy $F = F(\vec{\phi}_{j \in \{1,2,3\}})$ up to the fourth order in $\vec{\phi}_j$, and find the solution of $\vec{\phi}_j$ that minimizes $F$. Note that, due to time-reversal symmetry, only terms of even order in $\vec{\phi}_j$ are allowed. 
As such, there are only quadratic ($F_2$) and quartic terms ($F_4$) to be taken into account, $F = F_2+F_4$.
For simplicity, we only include the impact of SOC on the level of $F_2$ and take $F_4$ to be fully symmetric under all symmetries listed above. The general form of $F_4$ then reads
\begin{align}
F_4 =  c_1 \left( \sum_{j=1}^3\vec{\phi}^2_j \right)^2+c_2\sum_{j\neq j'}(\vec{\phi}_j\cdot\vec{\phi}_{j'})^2+c_3\sum_{j \neq j'}\vec{\phi}^2_j \vec{\phi}^2_{j'}\label{Eq. GLener}
\end{align}
with parameters $c_1, c_2, c_3$ (chosen to guarantee stability). Now, we write the symmetry-allowed $F_2$ for each case as
\begin{subequations}\begin{align}
\begin{split} \lambda_I = 0 , \  \lambda_R = 0 : F_2 &= a\sum_{j=1}^3 \vec{\phi}^2_j \end{split} \\
\begin{split}\lambda_I \neq 0, \  \lambda_R = 0: F_2 &= \sum_{j=1}^3 \left[a_1(\phi^z_j)^2 + a_2 (\vec{\varphi}^T_j \cdot \vec{\varphi}_j)\right]\end{split} \\
\begin{split}\lambda_I = 0, \  \lambda_R \neq 0 : F_2 &= \sum_{j=1}^3 \left[a_3 (\phi^z_j)^2 + \vec{\varphi}^T_j g_j \vec{\varphi}_j\right] \end{split} . 
\end{align}\label{GLener2}\end{subequations}
Here, we define the two component vectors $\vec{\varphi}_j = (\phi_j^x, \phi_j^y)^T$ and the $2\times2$ matrix $g_j = a_4 \mathbbm{1}_{2\times 2} \ + (R_{2\pi(j-1)/3} \vec{b})(R_{2\pi(j-1)/3} \vec{b})^T$, where $R_{2\pi(j-1)/3}$ is a 2x2 rotatiom matrix by an angle $2\pi(j-1)/3$.  We further introduced the real-valued parameters $(a, a_1, a_2, a_3, a_4)$ and the two-component real-valued vector $\vec{b}$. Without loss of generality, we fix the direction of $\vec{b}$ to be along the $x$ axis. From the symmetry point of view, Ising SOC sets $a_1 \neq a_2$, while Rashba SOC sets $a_3 \neq a_4$ and $\vec{b}\neq0$. Therefore, we parameterize the GL free-energy parameters $(a_1, a_2, a_3, a_4, \vec{b})$ with respect to the SOC parameters $(\lambda_I, \lambda_R)$ as $\vec{b} = \lambda_R \hat{e}_x$ and  
\begin{align}
\hspace{0cm}(a_1, a_2, a_3, a_4) = (a-\lambda^2_I, a+\lambda^2_I, a+\lambda^2_R, a-\lambda^2_R).\label{Eq. GLparameter}
\end{align}
Here, we absorbed the prefactors of the SOC parameters into the GL coefficients $(a_1, a_2, a_3, a_4)$ by redefining $\lambda_I$ and $\lambda_R$. The signs of the SOC corrections to $a_j$ in \equref{Eq. GLparameter} are chosen so that Ising and Rashba SOC favor out-of-plane and in-plane orders, respectively, in line with our HF numerics. Note that linear terms in \equref{Eq. GLparameter} are prohibited, as can be seen by performing $S_{\hat{x}\pi}$ and $S_{\hat{z}\pi}$ rotations which switch the signs of $\lambda_I$ and $\lambda_R$, respectively.

\begin{figure*}[t]
	\centering
	\includegraphics[width= 1.8\columnwidth, height= 0.4\columnwidth]{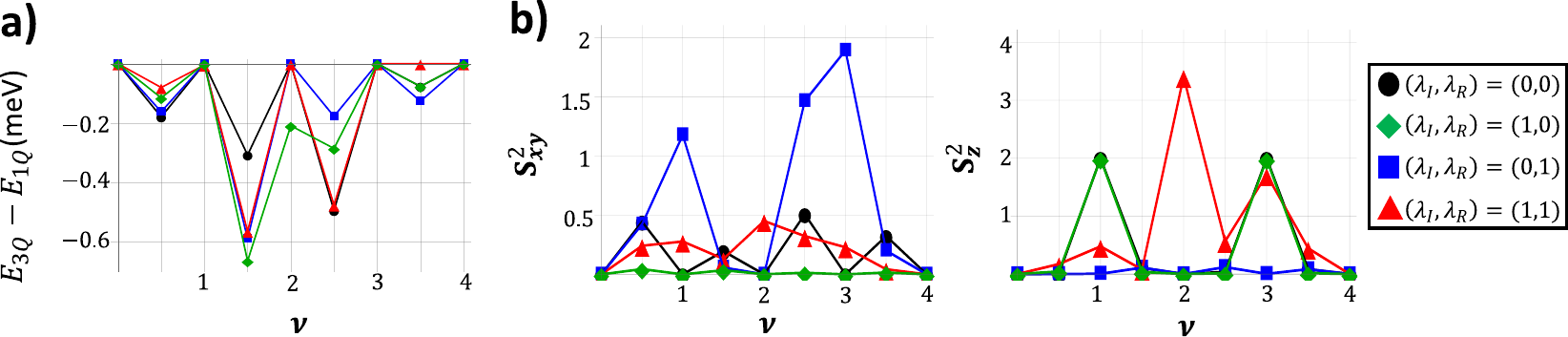}
	\caption{{\bf Single-band HF data of $3Q$ state}
    \justifying
    (\textbf{a}) HF energy difference between $1Q$ and $3Q$ states per moir\'e unit cell as a function of filling-factor $\nu$ with different SOC parameters. (\textbf{b}) Filling-factor dependence of in-plane $\textbf{S}^2_{xy}$ and out-of-plane $\textbf{S}^2_z$ spin expectation values with different SOC parameters. The SOC parameters are in units of meV. The system size is $12\times12$.
	}
	\label{Fig9:HFenergySDW3Q}
\end{figure*}

\subsection{Zero SOC}
We start with the GL analysis for zero SOC. Our main interest is the phase diagram with respect to the parameter set $(c_2,c_3)$ for fixed $(a, c_1)$, chosen as $(a, c_1)= (-1,1)$ for concreteness. We find the configurations $\vec{\phi}_j$ which minimize $F$ for each $(c_2, c_3)$. Figure~\ref{Fig14:GLanalysis}(a) shows the resulting phase diagram with respect to $(c_2, c_3)$ in the range $-1\leq c_2, c_3 \leq 1$ for zero SOC. We see that depending on the values of $(c_2, c_3)$, the TAF, clSDW$_{1Q}$  or clSDW$_{3Q}$ state can be ground state, with all phase boundaries corresponding to first order phase transitions. 

\subsection{Ising SOC}
Now, it is instructive to focus on how the Ising SOC affects the TAF state. To start from the TAF state, we first choose the parameter set $(c_1, c_2, c_3) = (1,1,-1)$. 
Figure~\ref{Fig14:GLanalysis}(b) shows the GL phase diagram for nonzero Ising SOC with the parameter set $(a, \lambda_I)$ in the range \ $-0.1\leq a\leq 0.8$. We find a $\textrm{TAF} \rightarrow \textrm{ncpSDW} \rightarrow \textrm{clSDW}_{3Q}$ continuous phase transition driven by Ising SOC, where the two types of $\textrm{SDW}_{3Q}$ states ($\vec{\phi}_1 = \pm\vec{\phi}_2 = \vec{\phi}_3$) are energetically degenerate. We expect that this degeneracy will be broken if the reduction of the symmetry group SO(3)$\rightarrow$ SO(2) due to Ising SOC is also considered in the fourth-order terms of the GL free-energy. 

\subsection{Rashba SOC}
We next discuss how the Rashba SOC affects the TAF state. Similar to the previous section, to start from the TAF state, we fix the parameters of the quartic terms to be $(c_1, c_2, c_3) = (1,1,-1)$. The phase diagram for nonzero Rashba SOC with different values of $(a , \lambda_R)$ in a range of $-0.1\leq a\leq 0.8$ is shown in Fig.\ \ref{Fig14:GLanalysis}(c). We clearly see the $\textrm{TAF} \rightarrow \textrm{ncpSDW} \rightarrow \textrm{cpSDW}$ continuous phase transition induced by Rashba SOC. Here, we point out again that we find two types of cpSDW states [see also Fig.~\ref{Fig14:GLanalysis}(d)] that are energetically degenerate in our GL model. We expect that this degeneracy will be broken if the modified symmetry ($C_{3z}, S_{\varphi} \rightarrow C_{3z} \otimes S_{\varphi}$) due to Rashba SOC is also considered in the fourth-order terms of the GL free-energy. 

\subsection{Broken time-reversal symmetry}\label{BTRS}
So far, we focused on the case where time-reversal symmetry is preserved before the onset of finite $\vec{\phi}_j$, such that the free-energy expansion is constrained by $\theta$.
However, for specific filling-factors (such as $\nu = 1/2$), time-reversal symmetry is already broken due to spontaneous valley polarization, allowing a third-order term $F_3 = c[ \vec{\phi}_1\cdot (\vec{\phi}_2\times\vec{\phi}_3)]$ to appear. The key changes in the phase diagrams when replacing the GL free-energy $F$ with $F+F_3$ for nonzero $c$ are the following (setting $c=-1$ in the numerical minimization for concreteness). First, we find that the qualitative features of the phase diagrams in Fig.~\ref{Fig14:GLanalysis}(a) and (b) are robust. For the phase diagram in Fig.\ \ref{Fig14:GLanalysis}(c), however, our GL analysis shows that the phase transition between ncpSDW and cpSDW actually becomes a smooth crossover, i.e., we find $\chi_{\textrm{GL}} > 0$, $m_{12} = -\frac{1}{2}, m_{23}= m_{31} = \pm\frac{1}{2}$ and hence formally a ncpSDW phase for all parameters in Fig.~\ref{Fig14:GLanalysis}(c).

\section{HF analysis in conduction bands: $3Q$ state}
\label{Sec:HF3Q}
\subsection{Energy comparison between $1Q$ and $3Q$ states}\label{SBH1Q}
With the intuition about possible $3Q$ solutions at hand, we return to a HF analysis and extend the variational space to include all three $M$-point momenta, $\mathcal{Q} = \{\Gamma, \  M_1, \  M_2, \  M_3\}$, to capture possible $3Q$ states. 

Similar to \secref{EnCompTI1Q}, we first consider the differences of the HF energy between the solution obtained from the $1Q$ and $3Q$ variational spaces, shown in Figure \ref{Fig9:HFenergySDW3Q}(a) for different filling factors and SOC parameters. We clearly see that regardless of the chosen SOC parameters, a $3Q$ state becomes energetically favored over a $1Q$ state for all half-integer filling factors, except at $\nu = 7/2$ and $(\lambda_I, \lambda_R) = (1,1) \ \textrm{meV}$, where the variational $3Q$ ansatz converges to a $1Q$ state. The presence of SOC still quantitatively affects the energy difference per moir\'e unit cell between $3Q$ and $1Q$ states. We also point out that for $(\lambda_I, \lambda_R) = (1,1) \ \textrm{meV}$ and $\nu=3/2$, a $3Q$ state becomes energetically favored over the TI state [see Fig. \ref{Fig4:HFTMBG}(a)]. For integer fillings, we see that our variational $3Q$ ansatz converges to a TI state, except at $\nu=1$ and $(\lambda_I, \lambda_R) = (0,1) \ \textrm{meV}$, where the $3Q$ HF calculation converges to a $1Q$ state [cf.~\figref{Fig4:HFTMBG}(a)]. For all considered fillings and different SOC parameters, the HF ground states are insulating; see \appref{BandGap}. 

\subsection{Spin configuration}
As before in \secref{SVPSs}, we study the correlator matrix $\vec{P}_{\textrm{3}Q}(\vec{k}) \equiv \vec{P}(\vec{k}, \mathcal{Q} = \{\Gamma, M_1, M_2, M_3\})$ averaged over momentum: $\tilde{\textbf{P}}_{\textrm{$3Q$}} \equiv \frac{1}{N_{3Q}}\sum_{\vec{k}\in \textrm{mBZ}_{3Q}}\vec{P}_{\textrm{$3Q$}}(\vec{k})$. The expectation value of general combinations of Pauli matrices $O$ can then be obtained as 
\begin{align}
\langle O\rangle_{3Q} = \textrm{Tr}[\tilde{\textbf{P}}_{\textrm{$3Q$}}O]\label{expectationSDW$_{3Q}$}.
\end{align}
In addition, we study how the SOC affects the spin configurations of the $3Q$ states by investigating the in-plane, $(\textbf{S}^2_{xy})_{3Q} = (\textbf{S}^2_x)_{3Q}+(\textbf{S}^2_y)_{3Q}$, and out-of-plane spin order, $(\textbf{S}^2_z)_{3Q}$. Here we use the definition
\begin{align}
(\mathbf{S}^2_{\alpha})_{3Q} = \frac{1}{4} \sum_{j,k, l}\langle s_{\alpha}\tau_j \lambda_{kl}
\rangle ^2 \label{eq:spinSDW$_{3Q}$},
\end{align}
where $\lambda_{kl}\equiv\sigma_k \otimes \sigma_l$ is used to parametrize a matrix basis on the four-dimensional space associated with $\{\Gamma, M_1, M_2, M_3\}$. We sum over all sixteen basis matrices in order to extract the spin components in all momentum sectors. 

Fig.~\ref{Fig9:HFenergySDW3Q}(b) shows the filling-factor dependence of the in- and out-of-plane components for different SOC parameters. Similar to  Fig.\ \ref{Fig5:HFTMBG2}(b), we observe that the in-plane spin order parameter $\mathbf{S}^2_{xy}$ is mostly suppressed throughout the entire filling-factor range for $(\lambda_I, \lambda_R) = (1,0) \ \textrm{meV}$ due to the spin-valley locking term from the Ising SOC Hamiltonian. Meanwhile, $\mathbf{S}^2_{z}$ is suppressed throughout the entire filling-factor range for $(\lambda_I, \lambda_R) = (0,1) \ \textrm{meV}$, resulting from the $s_x, s_y$ couplings in the Rashba SOC Hamiltonian. 

\begin{figure*}[t]
	\centering
	\includegraphics[width=\linewidth]{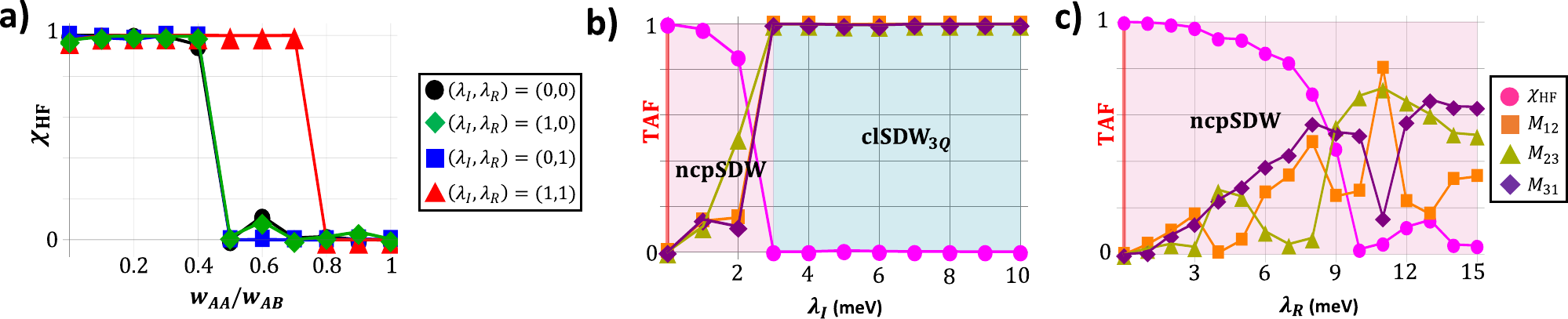}
	\caption{{\bf Spin chirality and inner product of spin magnetization for $3Q$ states}
    \justifying
    (\textbf{a}) Spin chirality $\chi_{\textrm{HF}}$ as a function of $w_{AA}/w_{AB}$ at filling-factor $\nu=1/2$ with fixed $w_{AB} = 110 \ \textrm{meV}$ and different SOC parameters. (\textbf{b}) Spin chirality $\chi_{\textrm{HF}}$ and scalar products $M_{12}, \ M_{23}, \ M_{31}$ as a function of Ising SOC $\lambda_I$ up to 10 meV for filling-factor $\nu = 1/2$, $w_{AA}/w_{AB}=0$ and $\lambda_R=0$.  (\textbf{c}) Same quantities as in (\textbf{b}) as a function of Rashba SOC $\lambda_R$ up to 15 meV for filling-factor $\nu = 1/2$, $w_{AA}/w_{AB}=0$ and $\lambda_I=0$. The SOC parameters are in units of meV. The system size is $12\times12$.
	}
	\label{Fig10:chi}
\end{figure*}

\subsection{Spin chirality and product}
\label{chiprod}
In this section, we focus on how the SOC parameters affect the HF analogues of spin chirality and spin scalar products defined in \equref{mAndChiDefinition}. To this end, we focus on a filling-factor of $\nu = 1/2$, where a valley polarized $3Q$ state becomes the HF ground state, and compare it to our GL analysis results from \secref{GLSection}. 

For a gauge-invariant, numerical evaluation within our HF theory, we first compute the net spin expectation value in real space $S_{a\beta l}(\vec{r})$ for given spin component $a$, sublattice $\beta$, layer $l$ and valley $\tau$ indices, which can be written as 
\begin{align}\begin{split}
&S_{a\beta l}(\vec{r}) = \frac{1}{N_{3Q}}\sum_{\vec{k}\in \textrm{mBZ}_{3Q}}\sum_{\substack{i,j \in\\\{0,1,2,3\}}}\sum_{\substack{\vec{G},\vec{G'}\\b_1,b_2}}
\langle c^{\dagger}_{\vec{k}+M_i b_1 \tau} c^\pdagger_{\vec{k}+M_j b_2 \tau} \rangle \end{split} \nonumber \\
\begin{split}& \ \ \ \ \ \ \ \ \ \ \ \times [u_{\vec{G}b_1\tau}(\vec{k}+M_i)]_{\beta ls} [s_{a}]_{ss'}[u_{\vec{G}'b_2\tau}(\vec{k}+M_j)]_{\beta l s'} \nonumber \end{split} \\
\begin{split}
& \ \ \ \ \ \ \ \ \ \ \ \times e^{-i(M_i-M_j+\vec{G}-\vec{G'})\cdot \vec{r}},
\label{spinexpreal}\end{split}\end{align}
with system size $N_{3Q}$, spin Pauli matrix vector $s_{a}$, $a\in\{0,1,2,3\}$ and $s_0\equiv\mathbbm{1}$, band indices $b_1, b_2$, three $M_1, M_2, M_3$ points and $M_0 \equiv\Gamma$. Here, for given momentum $\vec{k}, \vec{k'}\in\textrm{mBZ}_{3Q}$, we have already used the relation $\langle c^{\dagger}_{\vec{k}+M_i b_1 \tau} c_{\vec{k}'+M_j b_2 \tau} \rangle \sim \delta_{\vec k,\vec k'}$. In addition, for given moir\'e reciprocal vector $\vec{G}$, we define $u_{\vec{G}b_1\tau}(\vec{k})$ as the $\vec{G}$ component of BM wavefunction $u_{b_1\tau}(\vec{k})$ (see \secref{spTMBG}), which is written in sublattice $\beta$, layer $l$ and spin $s$ basis. We use four $AA$ sites $\vec{R}_0 = (0,0), \vec{R}_1 = L(0,1), \vec{R}_2 = L(\frac{\sqrt{3}}{2},\frac{1}{2}), \vec{R}_3 = L(\frac{\sqrt{3}}{2},\frac{3}{2})$ in the enlarged ($2\times2$) moiré unit cell, along with the three $M$ points $M_1=K_{\theta}(\sqrt{3}/2,0)$, $M_2=K_{\theta}(-\sqrt{3}/4,-3/4)$, $M_3=K_{\theta}(\sqrt{3}/4,-3/4)$. Here, $L = a_0/(2 \sin(\theta/2))$ is the moiré unit cell length, $a_0$ is the graphene lattice constant, and $\theta$ is the twist angle. The spin magnetization $\Phi_{ja\beta l} \ (j\in\{1,2,3\})$ and ferromagnetic contribution $\Phi_{0a\beta l}$ (not included in Eq.\ (\ref{SpinExpGL})) can then be extracted by solving the following four equations for $j \in \{0,1,2,3\}$
    \begin{align}
    \begin{split}
   S_{\alpha\beta l}(\vec{R}_j) = \Phi_{0a\beta l} + \sum_{i=1}^3\textrm{cos}(M_i\cdot\vec{R}_j) \Phi_{ja\beta l}
    \end{split}.
    \end{align}
For chosen $\beta$ and $l$, by writing the spin magnetization vector as $\vec{\Phi}_{j\beta l} =(\Phi_{jx\beta l}, \Phi_{jy\beta l},\Phi_{jz\beta l}) \ (j\in\{1,2,3\})$, we define the normalized scalar spin chirality $\chi_{\textrm{HF}}$ and product $M_{ij}$ in our HF analysis as 
\begin{align}
\chi_{\textrm{HF}} = \frac{|\vec{\Phi}_{1\beta l}\cdot(\vec{\Phi}_{2\beta l}\times \vec{\Phi}_{3\beta l})|}{|\vec{\Phi}_{1\beta l}||\vec{\Phi}_{2\beta l}||\vec{\Phi}_{3\beta l}|}, \ \ M_{ij} = \frac{|\vec{\Phi}_{i\beta l} \cdot \vec{\Phi}_{j\beta l}|}{|\vec{\Phi}_{i\beta l}||\vec{\Phi}_{j\beta l}|} \label{eq:chirality}.
\end{align}
Since we are interested in the effect of SOC which comes from the TMD on top of TMBG  (see Fig.\ \ref{Fig0:Setup}), which is why we show here the result for sublattice $\beta=A$ and layer $l=1$ which is closest to the TMD, in Fig.\ \ref{Fig10:chi}. We refer to  \figref{Fig14:layerlatticeextradata} in the supplement for how the choice of sublattice and layer degrees of freedom changes $\chi_{\textrm{HF}}$ and $M_{ij}$. 
We also note that the numerical values of $|\vec{\Phi}_{1\beta l}|, |\vec{\Phi}_{2\beta l}|$ and $|\vec{\Phi}_{3\beta l}|$ are found to be not too small for the parameters we study here. Consequently, the numerical evaluation of Eq.\ (\ref{eq:chirality}) is not unstable. 

We first show $\chi_{\textrm{HF}}$ as a function of $w_{AA}/w_{AB}$ with different SOC parameters in Fig.\ \ref{Fig10:chi}(a). Our HF data clearly show the suppression of $\chi_{\textrm{HF}}$ from $1$ to $0$ upon increasing the ratio $w_{AA}/w_{AB}$ up to around $0.5$ for zero SOC (except at $w_{AA}/w_{AB}=0.6$, where we see suppressed but finite $\chi_{\textrm{HF}}\sim 0.1$). Up to numerical precision, this qualitatively matches well with our GL phase diagram (see Fig.\ \ref{Fig14:GLanalysis}(a)), where $w_{AA}/w_{AB}$ will change the value of $(c_2, c_3)$ and we can extract from the comparison that it will induce the $\textrm{TAF} \rightarrow \textrm{clSDW}_{3Q}$ phase transition. 

When a small SOC is present, we see that $\chi_{\textrm{HF}}$ remains almost unchanged in the region $0\leq w_{AA}/w_{AB}\lesssim 0.4$ where $\chi_{\textrm{HF}}=1$ without SOC. In detail, for $(\lambda_I, \lambda_R) = (1,0), \ (0,1), \ (1,1)$ meV, $\chi_{\textrm{HF}}$ changes only by around 2\%, 0.1\%, 2\% compared to $\chi_{\textrm{HF}}=1$ without SOC, respectively.  We expect that $1$ meV of Ising or Rashba SOC strength is not large enough to clearly induce the $\textrm{TAF} \rightarrow \textrm{ncpSDW}$ phase transition. 

In addition, for the range $0.5\leq w_{AA}/w_{AB}\leq1$, where $\chi_{\textrm{HF}}$ without SOC is suppressed to approximately zero, our HF data show that $\chi_{\textrm{HF}}$ stays at zero, when only one form of SOC is present ($(\lambda_I, \lambda_R) = (1,0), \ (0,1)$ meV). From our GL analysis, we also checked that (not shown) Ising SOC leaves $\textrm{clSDW}_{3Q}$ as the ground state, while Rashba SOC induces a continuous $\textrm{clSDW}_{3Q}\rightarrow \textrm{ncpSDW} \ (m_{12}=-\frac{1}{2}, m_{23}=m_{31}=\pm\frac{1}{2})$ phase transition when time reversal-symmetry is broken due to spontaneous valley polarization and the extra GL free-energy term $F_3$ is present. 
Assuming that $\lambda_R=1 \ \textrm{meV}$ is not large enough to clearly induce a $\textrm{clSDW}_{3Q}\rightarrow \textrm{ncpSDW}$ phase transition within HF, our GL analysis and HF data agree qualitatively. Therefore, going back to the behavior as function of $w_{AA}/w_{AB}$, our HF data also show the $\textrm{TAF} \rightarrow \textrm{clSDW}_{3Q}$ phase transition up to numerical precision driven by tuning $w_{AA}/w_{AB}$ for $(\lambda_I, \lambda_R) = (1,0) , \ (0,1) \ \textrm{meV}$  as in the case without SOC.

Similar to the previous cases, we also clearly see the $\textrm{TAF} \rightarrow \textrm{clSDW}_{3Q}$ transition when both Ising and Rashba SOC are present ($(\lambda_I, \lambda_R) = (1,1)$ meV). However, in contrast to the previous cases, where the phase transition occured around $w_{AA}/w_{AB}\simeq 0.5$, we see that for $(\lambda_I, \lambda_R) = (1,1) \ \textrm{meV}$, the $\textrm{TAF} \rightarrow \textrm{clSDW}_{3Q}$ transition occurs around $w_{AA}/w_{AB}\simeq 0.8$. We expect this to happen, because Ising and Rashba SOC favor out-of-plane and in-plane spin order, respectively, such that their combination induces frustration, which reintroduces or extends the regime of non-coplanar order.  

Next, we compare our HF data with Fig.~\ref{Fig14:GLanalysis}(b) and (c) in more detail, which describes how the TAF state evolves when increasing Ising and Rashba SOC. To do so, we focus on $w_{AA}/w_{AB} =0$, which allows the TAF state to be the ground state when SOC is absent, and plot $\chi_{\textrm{HF}}, \  M_{12},\  M_{23},\ M_{31}$ as a function of $\lambda_I \ (\lambda_R=0)$ up to 10 meV and $\lambda_R \ (\lambda_I=0)$ up to 15 meV in Fig.~\ref{Fig10:chi}(b) and (c), respectively. We first clearly see the $\textrm{TAF} \rightarrow \textrm{ncpSDW} \rightarrow \textrm{clSDW}_{3Q}$ phase transition driven by tuning $\lambda_I$ in Fig.~\ref{Fig10:chi}(b), where $\chi_{\textrm{HF}}$ decreases from 1 to 0 and $M_{12}, M_{23}, M_{31}$ increase from 0 to 1. We also see the qualitative agreement between Fig.~\ref{Fig10:chi}(c) and Fig.~\ref{Fig14:GLanalysis}(c), which shows the $\textrm{TAF} \rightarrow \textrm{ncpSDW}$ phase transition driven by tuning $\lambda_R$. Since our HF data shows that the $3Q$ states for $\nu = 1/2$, $\lambda_I=0$ and $0\leq \lambda_R \leq 15 \ \textrm{meV}$ are valley polarized, we expect these signatures of Fig.~\ref{Fig10:chi}(c) to be connected to the 
$\textrm{ncpSDW} \rightarrow \textrm{cpSDW}$ crossover
induced by the extra GL free-energy term $F_3$. This term changes the phase transition between ncpSDW and cpSDW in Fig.~\ref{Fig14:GLanalysis}(c) to a smooth crossover (see \secref{BTRS}). We point out that the behavior of $\chi_{\text{HF}}$ and $M_{ij}$ with $\lambda_R$ depends significantly on which layer and sublattice we consider: as detailed in \appref{OtherSublatticesLayers}, the transition and crossover mentioned above is most clearly visible in the sublattice and layer we chose in \figref{Fig10:chi}(c), while the change is much less prominent or even absent for the other choices. A potential reason for these differences lies in the explicit sublattice dependence of Rashba SOC together with the effective layer dependence from the distance to the topmost TMD layer. Interestingly, in contrast, when Ising SOC is only present, the transition is clearly visible not only in both sublattices but also in the layers further away from the TMD.

Finally, as already mentioned above, in the presence of SOC, a spin modulation is expected to also induce a finite charge modulation based on symmetry. To demonstrate that this is also the case within our HF numerics, we computed (not shown) the following charge-modulation order parameter $\Delta_{ij}\equiv |\sum_{\beta \in\{A,B\},  l\in\{1,2,3\}}(S_{0\beta l}(\vec{R}_i)-S_{0\beta l}(\vec{R}_j))|$, for two neighboring $AA$ sites $\vec{R}_i \ \textrm{and} \ \vec{R}_j$. We checked that for the TAF state, $\Delta_{ij}$ is numerically zero, 
while for the other states ($\textrm{clSDW}_{1Q}, \ \textrm{clSDW}_{3Q}, \ \textrm{ncpSDW}$), $\Delta_{ij}$ becomes finite.

\section{Conclusion}
\label{Sec:Conclusion}
We performed self-consistent HF calculations for the conduction bands of TMBG for all integer and half-integer fillings $0< \nu <4$ with different SOC parameters, allowing the HF solutions to also capture translational-symmetry-broken states on the moir\'e scale. We found that, when SOC is absent, $1Q$ and $3Q$ states are the HF ground states for all half-integer fillings, whereas TI states are the HF ground states for all integer fillings. This energetic competition between TI and translational-symmetry-broken states remains mostly robust upon the addition of proximitized SOC. However, the spin and valley nature of the ground states is very sensitive to the presence of SOC, even to small SOC parameters on the order of $\sim1 \ \textrm{meV}$. Generally, we observe that when only Ising SOC is present, in-plane spin order becomes suppressed and SVP is induced due to the spin-valley locking term of the Ising SOC Hamiltonian. In contrast, when only Rashba SOC is present, out-of-plane spin order and SVP are suppressed due to the $s_x, s_y$ terms from the Rashba SOC Hamiltonian. In addition, our combined GL and HF analysis for different 3Q states showed that Ising SOC drives a $\textrm{TAF} \rightarrow \textrm{ncpSDW} \rightarrow\textrm{clSDW}_{3Q}$ phase transition, whereas Rashba SOC drives a $\textrm{TAF} \rightarrow \textrm{ncpSDW} \rightarrow\textrm{cpSDW}$ phase transition. We identified these states by their spin chirality $\chi_{\textrm{HF}}$ and the angle between spin magnetizations, quantified by the scalar products $M_{12}, M_{23}, M_{31}$ [see \equref{eq:chirality} for their definition]. 
Finally, we additionally performed HF calculations taking the valence bands into account, see App. \ref{WithValenceBands}. They confirm our findings from the HF calculations for the conduction band for fillings $0<\nu\leq4$. In addition, we observe a strong band hybridization, which, in particular, induces a non-zero indirect band gap between conduction and valence bands for $\nu=0$. 

Our results clearly demonstrate the interesting effect of SOC on the correlated ground states of TMBG for all integer and half-integer fillings. The spin nature of translational-symmetry-broken states at half-integer fillings varies significantly, ranging from collinear or coplanar states to non-coplanar states with finite spin chirality. In particular, when both Rashba and Ising SOC are present, as is generally the case in an inversion-asymmetric moiré structure like SOC-TMBG, the resulting frustration can change the ground state from a collinear state to a chiral, non-coplanar state.

The effect of induced SOC on TMBG is therefore most pronounced away from integer fillings. 
Complete valley polarization at integer fillings in TMBG can be observed via the quantum anomalous Hall effect \cite{polshyn2020electrical}.  
Furthermore, the valley Chern numbers imply that the partially valley-polarized states at half-integer fillings also display a non-zero anomalous Hall effect. Such anomalous Hall features indicating a Chern number $C=1$ were already observed in TMBG without SOC at $\nu=3/2$ and $\nu=7/2$ \cite{polshyn2020electrical, xu2021tunable, chen2021electrically, he2021competing}. Similar states were recently suggested for other graphene structures \cite{dong2024anomalous,tan2024parent}. Our prediction adds the complexity of non-coplanar spin order to such topologically nontrivial charge-ordered states. As we showed, the non-coplanar SDW is intertwined with charge order, where the inter-site density modulations can be observed using scanning tunneling microscopy \cite{Nuckolls.2024}. In addition, the spin order can, in principle, be probed using local magnetometry techniques, such as SQUID-on-tip measurements or a spin-resolved scanning tunneling microscope. Therefore, detection of the exotic phases of SOC-TMBG predicted in this paper requires a combination of transport measurements and local probes of spin and charge.

\begin{acknowledgments}
We thank Sida Tian, Achim Rosch, Robin Scholle and, Yves.\ H.\ Kwan for fruitful discussion. J.Y.P thanks the Kwanjeong Educational Foundation for support. L.C. was funded by the European Union (ERC-2023-STG, Project 101115758 - QuantEmerge). M.S.S.~acknowledges support from the European Union (ERC-2021-STG, Project 101040651---SuperCorr). JHS received financial support from the DFG through SPP 2244. Views and opinions expressed are those of the authors only and do not necessarily reflect those of the European Union or the European Research Council Executive Agency. Neither the European Union nor the granting authority can be held responsible for them.

\end{acknowledgments}

\bibliography{draft_Refs}

@article{dong2024anomalous,
  title={Anomalous Hall crystals in rhombohedral multilayer graphene. I. Interaction-driven Chern bands and fractional quantum Hall states at zero magnetic field},
  author={Dong, Junkai and Wang, Taige and Wang, Tianle and Soejima, Tomohiro and Zaletel, Michael P and Vishwanath, Ashvin and Parker, Daniel E},
  journal={Physical Review Letters},
  volume={133},
  number={20},
  pages={206503},
  year={2024},
  url={https://journals.aps.org/prl/abstract/10.1103/PhysRevLett.133.206503},
  publisher={APS}
}

@article{tan2024parent,
  title={Parent Berry curvature and the ideal anomalous Hall crystal},
  author={Tan, Tixuan and Devakul, Trithep},
  journal={Physical Review X},
  volume={14},
  number={4},
  url={https://journals.aps.org/prx/abstract/10.1103/PhysRevX.14.041040},
  pages={041040},
  year={2024},
  publisher={APS}
}

@article{Nuckolls.2024, 
year = {2024}, 
title = {{A microscopic perspective on moiré materials}}, 
author = {Nuckolls, Kevin P. and Yazdani, Ali}, 
journal = {Nature Reviews Materials}, 
doi = {10.1038/s41578-024-00682-1}, 
eprint = {2404.08044}, 
pages = {460--480}, 
number = {7}, 
volume = {9}
}

@article{park2023network,
  title={Network of chiral one-dimensional channels and localized states emerging in a moir{\'e} system},
url = {https://iopscience.iop.org/article/10.1088/2053-1583/acdd82/meta},
  author={Park, Jeyong and Gresista, Lasse and Trebst, Simon and Rosch, Achim and Park, Jinhong},
  journal={2D Materials},
  volume={10},
  number={3},
  pages={035033},
  year={2023},
  publisher={IOP Publishing}
}

@article{bistritzer2011moire,
  title={Moir{\'e} bands in twisted double-layer graphene},
url = {https://www.pnas.org/doi/abs/10.1073/pnas.1108174108},
  author={Bistritzer, Rafi and MacDonald, Allan H},
  journal={Proceedings of the National Academy of Sciences},
  volume={108},
  number={30},
  pages={12233--12237},
  year={2011},
  publisher={National Acad Sciences}
}

@article{wang2026family,
  title={Family of High-Chern-Number Orbital Magnets in Twisted Rhombohedral Graphene},
url={https://arxiv.org/abs/2601.01087},
  author={Wang, Xirui and Ben{\'\i}tez, L Antonio and Phong, Vo Tien and Chu, Wai In and Watanabe, Kenji and Taniguchi, Takashi and Lewandowski, Cyprian and Jarillo-Herrero, Pablo},
  journal={arXiv preprint arXiv:2601.01087},
  year={2026}
}

@article{zhang2025entangled,
  title={Entangled Moire Chern Insulator in Rhombohedral Graphene},
  author={Zhang, Zaizhe and Chen, Xi and Watanabe, Kenji and Taniguchi, Takashi and Song, Zhida and Lu, Xiaobo},
  journal={arXiv preprint arXiv:2512.21609},
url={https://arxiv.org/abs/2512.21609},
  year={2025}
}

@article{yu2023magic,
  title={Magic-angle twisted symmetric trilayer graphene as a topological heavy-fermion problem},
  author={Yu, Jiabin and Xie, Ming and Bernevig, B Andrei and Das Sarma, Sankar},
  journal={Physical Review B},
  volume={108},
  number={3},
url={https://journals.aps.org/prb/abstract/10.1103/PhysRevB.108.035129},
  pages={035129},
  year={2023},
  publisher={APS}
}

@article{huang2025angle,
  title={Angle-tuned Gross-Neveu quantum criticality in twisted bilayer graphene},
  author={Huang, Cheng and Parthenios, Nikolaos and Ulybyshev, Maksim and Zhang, Xu and Assaad, Fakher F and Classen, Laura and Meng, Zi Yang},
  journal={Nature Communications},
url={https://www.nature.com/articles/s41467-025-62461-y},
  volume={16},
  number={1},
  pages={7176},
  year={2025},
  publisher={Nature Publishing Group UK London}
}

@article{sun2023determining,
  title={Determining spin-orbit coupling in graphene by quasiparticle interference imaging},
  author={Sun, Lihuan and Rademaker, Louk and Mauro, Diego and Scarfato, Alessandro and P{\'a}sztor, {\'A}rp{\'a}d and Guti{\'e}rrez-Lezama, Ignacio and Wang, Zhe and Martinez-Castro, Jose and Morpurgo, Alberto F and Renner, Christoph},
  journal={Nature communications},
  volume={14},
url={https://www.nature.com/articles/s41467-023-39453-x},
  number={1},
  pages={3771},
  year={2023},
  publisher={Nature Publishing Group UK London}
}

@article{huang2024evolution,
  title={Evolution from quantum anomalous Hall insulator to heavy-fermion semimetal in magic-angle twisted bilayer graphene},
  author={Huang, Cheng and Zhang, Xu and Pan, Gaopei and Li, Heqiu and Sun, Kai and Dai, Xi and Meng, Zi Yang},
  journal={Physical Review B},
  volume={109},
  number={12},
url={https://journals.aps.org/prb/abstract/10.1103/PhysRevB.109.125404},
  pages={125404},
  year={2024},
  publisher={APS}
}

@article{polshyn2020electrical,
  title={Electrical switching of magnetic order in an orbital Chern insulator},
  author={Polshyn, Hryhoriy and Zhu, Jihang and Kumar, Manish A and Zhang, Yuxuan and Yang, Fangyuan and Tschirhart, Charles L and Serlin, Marec and Watanabe, Kenji and Taniguchi, Takashi and MacDonald, Allan H and others},
url={https://www.nature.com/articles/s41586-020-2963-8},
  journal={Nature},
  volume={588},
  number={7836},
  pages={66--70},
  year={2020},
  publisher={Nature Publishing Group UK London}
}

@article{mccann2013electronic,
  title={The electronic properties of bilayer graphene},
url={https://iopscience.iop.org/article/10.1088/0034-4885/76/5/056503/meta?casa_token=vrkM7pKGq3MAAAAA:zsRI92d-m4j-x9uFxKohj8-67GzY8kd_lz256Lr--OJbwx9-Kl9yhE8OtsQZSU6RvoKGtVdwovhAr0TuUCIlIxU6HXQ},
  author={McCann, Edward and Koshino, Mikito},
  journal={Reports on Progress in physics},
  volume={76},
  number={5},
  pages={056503},
  year={2013},
  publisher={IOP Publishing}
}

@article{park2020gate,
  title={Gate-tunable topological flat bands in twisted monolayer-bilayer graphene},
url={https://journals.aps.org/prb/abstract/10.1103/PhysRevB.102.035411},
  author={Park, Youngju and Chittari, Bheema Lingam and Jung, Jeil},
  journal={Phys. Rev. B},
  volume={102},
  number={3},
  pages={035411},
  year={2020},
  publisher={APS}
}

@article{lin2022spin,
  title={Spin-orbit--driven ferromagnetism at half moir{\'e} filling in magic-angle twisted bilayer graphene},
url={https://www.science.org/doi/full/10.1126/science.abh2889},
  author={Lin, Jiang-Xiazi and Zhang, Ya-Hui and Morissette, Erin and Wang, Zhi and Liu, Song and Rhodes, Daniel and Watanabe, K and Taniguchi, T and Hone, James and Li, JIA},
  journal={Science},
  volume={375},
  number={6579},
  pages={437--441},
  year={2022},
  publisher={American Association for the Advancement of Science}
}

@article{tan2024topological,
  title={Topological phases, van Hove singularities, and spin texture in magic-angle twisted bilayer graphene in the presence of proximity-induced spin-orbit couplings},
url = {https://journals.aps.org/prb/abstract/10.1103/PhysRevB.110.165406},
  author={Tan, Yuting and Chou, Yang-Zhi and Wu, Fengcheng and Das Sarma, Sankar},
  journal={Phys. Rev. B},
  volume={110},
  number={16},
  pages={165406},
  year={2024},
  publisher={APS}
}

@article{rademaker2020topological,
  title={Topological flat bands and correlated states in twisted monolayer-bilayer graphene},
url={https://journals.aps.org/prresearch/abstract/10.1103/PhysRevResearch.2.033150},
  author={Rademaker, Louk and Protopopov, Ivan V and Abanin, Dmitry A},
  journal={Phys. Rev. Res.},
  volume={2},
  number={3},
  pages={033150},
  year={2020},
  publisher={APS}
}

@article{chen2021electrically,
  title={Electrically tunable correlated and topological states in twisted monolayer--bilayer graphene},
  author={Chen, Shaowen and He, Minhao and Zhang, Ya-Hui and Hsieh, Valerie and Fei, Zaiyao and Watanabe, Kenji and Taniguchi, Takashi and Cobden, David H and Xu, Xiaodong and Dean, Cory R and others},
url ={https://www.nature.com/articles/s41567-020-01062-6},
  journal={Nature Physics},
  volume={17},
  number={3},
  pages={374--380},
  year={2021},
  publisher={Nature Publishing Group UK London}
}

@article{wang2024chern,
  title={Chern-Textured Exciton Insulators with Valley Spiral Order in Moir$\backslash$'e Materials},
url={https://arxiv.org/abs/2406.15342},
  author={Wang, Ziwei and Kwan, Yves H and Wagner, Glenn and Simon, Steven H and Bultinck, Nick and Parameswaran, SA},
  journal={arXiv preprint arXiv:2406.15342},
  year={2024}
}

@article{bultinck2020ground,
  title={Ground state and hidden symmetry of magic-angle graphene at even integer filling},
url={https://journals.aps.org/prx/abstract/10.1103/PhysRevX.10.031034},
  author={Bultinck, Nick and Khalaf, Eslam and Liu, Shang and Chatterjee, Shubhayu and Vishwanath, Ashvin and Zaletel, Michael P},
  journal={Phys. Rev. X},
  volume={10},
  number={3},
  pages={031034},
  year={2020},
  publisher={APS}
}

@article{ingham2025moir,
  title={Moir\'e $\textrm{M}$-valley bilayers: quasi-one-dimensional physics, unconventional spin textures and twisted van Hove singularities},
url={https://arxiv.org/abs/2503.11754},
  author={Ingham, Julian and Scheurer, Mathias S and Scammell, Harley D},
  journal={arXiv preprint arXiv:2503.11754},
  year={2025}
}

@article{fukui2005chern,
  title={Chern numbers in discretized Brillouin zone: efficient method of computing (spin) Hall conductances},
url={https://journals.jps.jp/doi/abs/10.1143/JPSJ.74.1674},
  author={Fukui, Takahiro and Hatsugai, Yasuhiro and Suzuki, Hiroshi},
  journal={Journal of the Physical Society of Japan},
  volume={74},
  number={6},
  pages={1674--1677},
  year={2005},
  publisher={The Physical Society of Japan}
}

@article{zhang2019twisted,
  title={Twisted bilayer graphene aligned with hexagonal boron nitride: Anomalous Hall effect and a lattice model},
  author={Zhang, Ya-Hui and Mao, Dan and Senthil, Th},
  journal={Physical Review Research},
url={https://journals.aps.org/prresearch/abstract/10.1103/PhysRevResearch.1.033126},
  volume={1},
  number={3},
  pages={033126},
  year={2019},
  publisher={APS}
}

@article{hawashin2025relativistic,
  title={Relativistic Mott transition, high-order van Hove singularity, and mean-field phase diagram of twisted double bilayer $\textrm{WSe}_2$},
  author={Hawashin, Bilal and Kleeschulte, Julian and Kurz, David and Scherer, Michael M},
url={https://arxiv.org/abs/2509.09398},
  journal={arXiv preprint arXiv:2509.09398},
  year={2025}
}

@article{dong2025observation,
  title={Observation of integer and fractional Chern insulators in high Chern number flatbands},
  author={Dong, Jingwei and Liu, Le and Zhu, Jundong and Pan, Zitian and Hong, Yu and Jia, Zhengnan and Watanabe, Kenji and Taniguchi, Takashi and Du, Luojun and Shi, Dongxia and others},
url={https://arxiv.org/abs/2507.09908},
  journal={arXiv preprint arXiv:2507.09908},
  year={2025}
}

@article{wang2025moir,
  title={Moir$\backslash$'e dependent Chern insulators in twisted crystalline flatbands},
  author={Wang, Wenxuan and Wang, Yijie and Zhang, Zaizhe and Huo, Zihao and Zhou, Gengdong and Watanabe, Kenji and Taniguchi, Takashi and Xie, XC and Liu, Kaihui and Song, Zhida and others},
url={https://arxiv.org/abs/2507.10875},
  journal={arXiv preprint arXiv:2507.10875},
  year={2025}
}

@article{su2025moire,
  title={Moir{\'e}-driven topological electronic crystals in twisted graphene},
  author={Su, Ruiheng and Waters, Dacen and Zhou, Boran and Watanabe, Kenji and Taniguchi, Takashi and Zhang, Ya-Hui and Yankowitz, Matthew and Folk, Joshua},
  journal={Nature},
url={https://www.nature.com/articles/s41586-024-08239-6},
  volume={637},
  number={8048},
  pages={1084--1089},
  year={2025},
  publisher={Nature Publishing Group UK London}
}

@article{he2021competing,
  title={Competing correlated states and abundant orbital magnetism in twisted monolayer-bilayer graphene},
  author={He, Minhao and Zhang, Ya-Hui and Li, Yuhao and Fei, Zaiyao and Watanabe, Kenji and Taniguchi, Takashi and Xu, Xiaodong and Yankowitz, Matthew},
url={https://www.nature.com/articles/s41467-021-25044-1},
  journal={Nature Communications},
  volume={12},
  number={1},
  pages={4727},
  year={2021},
  publisher={Nature Publishing Group UK London}
}

@article{jiang2019charge,
  title={Charge order and broken rotational symmetry in magic-angle twisted bilayer graphene},
  author={Jiang, Yuhang and Lai, Xinyuan and Watanabe, Kenji and Taniguchi, Takashi and Haule, Kristjan and Mao, Jinhai and Andrei, Eva Y},
url={https://www.nature.com/articles/s41586-019-1460-4},
  journal={Nature},
  volume={573},
  number={7772},
  pages={91--95},
  year={2019},
  publisher={Nature Publishing Group UK London}
}

@article{su2023superconductivity,
  title={Superconductivity in twisted double bilayer graphene stabilized by WSe2},
  author={Su, Ruiheng and Kuiri, Manabendra and Watanabe, Kenji and Taniguchi, Takashi and Folk, Joshua},
url={https://www.nature.com/articles/s41563-023-01653-7},
  journal={Nature Materials},
  volume={22},
  number={11},
  pages={1332--1337},
  year={2023},
  publisher={Nature Publishing Group UK London}
}

@article{song2025fractional,
  title={Fractional Chern Insulators Transition in Non-ideal Flat Bands of Twisted Mono-bilayer Graphene},
  author={Song, Moru and Chang, Kai},
  journal={arXiv preprint arXiv:2511.12231},
url={https://arxiv.org/abs/2511.12231},
  year={2025}
}

@article{liu2025diverse,
  title={Diverse high-Chern-number quantum anomalous Hall insulators in twisted rhombohedral graphene},
  author={Liu, Naitian and Chen, Zhangyuan and Ding, Jing and Zhou, Wenqiang and Xiang, Hanxiao and Fang, Xinjie and Wu, Linfeng and Zhan, Xiaowan and Zhang, Le and Chen, Qianmei and others},
url={https://arxiv.org/abs/2507.11347},
  journal={arXiv preprint arXiv:2507.11347},
  year={2025}
}

@article{shi2022heavy,
  title={Heavy-fermion representation for twisted bilayer graphene systems},
  author={Shi, Hao and Dai, Xi},
  journal={Physical Review B},
url={https://journals.aps.org/prb/abstract/10.1103/PhysRevB.106.245129},
  volume={106},
  number={24},
  pages={245129},
  year={2022},
  publisher={APS}
}

@article{cao2018unconventional,
  title={Unconventional superconductivity in magic-angle graphene superlattices},
  author={Cao, Yuan and Fatemi, Valla and Fang, Shiang and Watanabe, Kenji and Taniguchi, Takashi and Kaxiras, Efthimios and Jarillo-Herrero, Pablo},
url={https://www.nature.com/articles/nature26160},
  journal={Nature},
  volume={556},
  number={7699},
  pages={43--50},
  year={2018},
  publisher={Nature Publishing Group UK London}
}

@article{po2018origin,
  title={Origin of Mott insulating behavior and superconductivity in twisted bilayer graphene},
url={https://journals.aps.org/prx/abstract/10.1103/PhysRevX.8.031089},
  author={Po, Hoi Chun and Zou, Liujun and Vishwanath, Ashvin and Senthil, T},
  journal={Phys. Rev. X},
  volume={8},
  number={3},
  pages={031089},
  year={2018},
  publisher={APS}
}

@article{chichinadze2020nematic,
  title={Nematic superconductivity in twisted bilayer graphene},
url={https://journals.aps.org/prb/abstract/10.1103/PhysRevB.101.224513},
  author={Chichinadze, Dmitry V and Classen, Laura and Chubukov, Andrey V},
  journal={Phys. Rev. B},
  volume={101},
  number={22},
  pages={224513},
  year={2020},
  publisher={APS}
}

@article{classen2019competing,
  title={Competing phases of interacting electrons on triangular lattices in moir{\'e} heterostructures},
url={https://journals.aps.org/prb/abstract/10.1103/PhysRevB.99.195120},
  author={Classen, Laura and Honerkamp, Carsten and Scherer, Michael M},
  journal={Phys. Rev. B},
  volume={99},
  number={19},
  pages={195120},
  year={2019},
  publisher={APS}
}

@article{PergeSOC,
	author = {Arora, Harpreet Singh and Polski, Robert and Zhang, Yiran and Thomson, Alex and Choi, Youngjoon and Kim, Hyunjin and Lin, Zhong and Wilson, Ilham Zaky and Xu, Xiaodong and Chu, Jiun-Haw and Watanabe, Kenji and Taniguchi, Takashi and Alicea, Jason and Nadj-Perge, Stevan},
	date = {2020/07/01},
	doi = {10.1038/s41586-020-2473-8},
	id = {Arora2020},
	isbn = {1476-4687},
	journal = {Nature},
	number = {7816},
	pages = {379--384},
	title = {Superconductivity in metallic twisted bilayer graphene stabilized by WSe2},
	url = {https://doi.org/10.1038/s41586-020-2473-8},
	volume = {583},
	year = {2020}}

@article{kim2022evidence,
  title={Evidence for unconventional superconductivity in twisted trilayer graphene},
url={https://www.nature.com/articles/s41586-022-04715-z},
  author={Kim, Hyunjin and Choi, Youngjoon and Lewandowski, Cyprian and Thomson, Alex and Zhang, Yiran and Polski, Robert and Watanabe, Kenji and Taniguchi, Takashi and Alicea, Jason and Nadj-Perge, Stevan},
  journal={Nature},
  volume={606},
  number={7914},
  pages={494--500},
  year={2022},
  publisher={Nature Publishing Group UK London}
}

@article{park2021tunable,
  title={Tunable strongly coupled superconductivity in magic-angle twisted trilayer graphene},
  author={Park, Jeong Min and Cao, Yuan and Watanabe, Kenji and Taniguchi, Takashi and Jarillo-Herrero, Pablo},
url={https://www.nature.com/articles/s41586-021-03192-0},
  journal={Nature},
  volume={590},
  number={7845},
  pages={249--255},
  year={2021},
  publisher={Nature Publishing Group UK London}
}

@article{xu2021tunable,
  title={Tunable van Hove singularities and correlated states in twisted monolayer--bilayer graphene},
url={https://www.nature.com/articles/s41567-021-01172-9},
  author={Xu, Shuigang and Al Ezzi, Mohammed M and Balakrishnan, Nilanthy and Garcia-Ruiz, Aitor and Tsim, Bonnie and Mullan, Ciaran and Barrier, Julien and Xin, Na and Piot, Benjamin A and Taniguchi, Takashi and others},
  journal={Nature Physics},
  volume={17},
  number={5},
  pages={619--626},
  year={2021},
  publisher={Nature Publishing Group UK London}
}

@article{scammell2023tunable,
  title={Tunable superconductivity and M{\"o}bius Fermi surfaces in an inversion-symmetric twisted van der Waals heterostructure},
url={https://journals.aps.org/prl/abstract/10.1103/PhysRevLett.130.066001},
  author={Scammell, Harley D and Scheurer, Mathias S},
  journal={Phys. Rev. Lett.},
  volume={130},
  number={6},
  pages={066001},
  year={2023},
  publisher={APS}
}

@article{chou2024topological,
  title={Topological flat bands, valley polarization, and interband superconductivity in magic-angle twisted bilayer graphene with proximitized spin-orbit couplings},
url={https://journals.aps.org/prb/abstract/10.1103/PhysRevB.110.L041108},
  author={Chou, Yang-Zhi and Tan, Yuting and Wu, Fengcheng and Das Sarma, Sankar},
  journal={Phys. Rev. B},
  volume={110},
  number={4},
  pages={L041108},
  year={2024},
  publisher={APS}
}

@article{kennes2021moire,
  title={Moir{\'e} heterostructures as a condensed-matter quantum simulator},
url={https://www.nature.com/articles/s41567-020-01154-3},
  author={Kennes, Dante M and Claassen, Martin and Xian, Lede and Georges, Antoine and Millis, Andrew J and Hone, James and Dean, Cory R and Basov, DN and Pasupathy, Abhay N and Rubio, Angel},
  journal={Nature Physics},
  volume={17},
  number={2},
  pages={155--163},
  year={2021},
  publisher={Nature Publishing Group UK London}
}

@article{efimkin2018helical,
  title={Helical network model for twisted bilayer graphene},
url={https://journals.aps.org/prb/abstract/10.1103/PhysRevB.98.035404},
  author={Efimkin, Dmitry K and MacDonald, Allan H},
  journal={Phys. Rev. B},
  volume={98},
  number={3},
  pages={035404},
  year={2018},
  publisher={APS}
}

@article{song2022magic,
  title={Magic-angle twisted bilayer graphene as a topological heavy fermion problem},
url={https://journals.aps.org/prl/abstract/10.1103/PhysRevLett.129.047601},
  author={Song, Zhi-Da and Bernevig, B Andrei},
  journal={Phys. Rev. Lett.},
  volume={129},
  number={4},
  pages={047601},
  year={2022},
  publisher={APS}
}

@article{gmitra2016trivial,
  title={Trivial and inverted Dirac bands and the emergence of quantum spin Hall states in graphene on transition-metal dichalcogenides},
url={https://journals.aps.org/prb/abstract/10.1103/PhysRevB.93.155104},
  author={Gmitra, Martin and Kochan, Denis and H{\"o}gl, Petra and Fabian, Jaroslav},
  journal={Phys. Rev. B},
  volume={93},
  number={15},
  pages={155104},
  year={2016},
  publisher={APS}
}

@article{gmitra2015graphene,
  title={Graphene on transition-metal dichalcogenides: A platform for proximity spin-orbit physics and optospintronics},
url={https://journals.aps.org/prb/abstract/10.1103/PhysRevB.92.155403},
  author={Gmitra, Martin and Fabian, Jaroslav},
  journal={Phys. Rev. B},
  volume={92},
  number={15},
  pages={155403},
  year={2015},
  publisher={APS}
}

@article{wang2016origin,
  title={Origin and magnitude of ‘designer’spin-orbit interaction in graphene on semiconducting transition metal dichalcogenides},
url={https://journals.aps.org/prx/abstract/10.1103/PhysRevX.6.041020},
  author={Wang, Zhe and Ki, Dong-Keun and Khoo, Jun Yong and Mauro, Diego and Berger, Helmuth and Levitov, Leonid S and Morpurgo, Alberto F},
  journal={Phys. Rev. X},
  volume={6},
  number={4},
  pages={041020},
  year={2016},
  publisher={APS}
}

@article{li2019twist,
  title={Twist-angle dependence of the proximity spin-orbit coupling in graphene on transition-metal dichalcogenides},
url={https://journals.aps.org/prb/abstract/10.1103/PhysRevB.99.075438},
  author={Li, Yang and Koshino, Mikito},
  journal={Phys. Rev. B},
  volume={99},
  number={7},
  pages={075438},
  year={2019},
  publisher={APS}
}

@article{PhysRevB.100.085412,
  title = {Induced spin-orbit coupling in twisted graphene--transition metal dichalcogenide heterobilayers: Twistronics meets spintronics},
  author = {David, Alessandro and Rakyta, P\'eter and Korm\'anyos, Andor and Burkard, Guido},
  journal = {Phys. Rev. B},
  volume = {100},
  issue = {8},
  pages = {085412},
  numpages = {15},
  year = {2019},
  month = {Aug},
  publisher = {American Physical Society},
  doi = {10.1103/PhysRevB.100.085412},
  url = {https://link.aps.org/doi/10.1103/PhysRevB.100.085412}
}

@article{fulop2021boosting,
  title={Boosting proximity spin--orbit coupling in graphene/WSe2 heterostructures via hydrostatic pressure},
url={https://www.nature.com/articles/s41699-021-00262-9},
  author={F{\"u}l{\"o}p, B{\'a}lint and M{\'a}rffy, Albin and Zihlmann, Simon and Gmitra, Martin and T{\'o}v{\'a}ri, Endre and Szentp{\'e}teri, B{\'a}lint and Kedves, M{\'a}t{\'e} and Watanabe, Kenji and Taniguchi, Takashi and Fabian, Jaroslav and others},
  journal={npj 2D Materials and Applications},
  volume={5},
  number={1},
  pages={82},
  year={2021},
  publisher={Nature Publishing Group UK London}
}

@article{PhysRevLett.127.026401,
  title = {Emulating Heavy Fermions in Twisted Trilayer Graphene},
  author = {Ramires, Aline and Lado, Jose L.},
  journal = {Phys. Rev. Lett.},
  volume = {127},
  issue = {2},
  pages = {026401},
  numpages = {6},
  year = {2021},
  month = {Jul},
  publisher = {American Physical Society},
  doi = {10.1103/PhysRevLett.127.026401},
  url = {https://link.aps.org/doi/10.1103/PhysRevLett.127.026401}
}

@article{parker2021field,
  title={Field-tuned and zero-field fractional Chern insulators in magic angle graphene},
url={https://arxiv.org/abs/2112.13837},
  author={Parker, Daniel and Ledwith, Patrick and Khalaf, Eslam and Soejima, Tomohiro and Hauschild, Johannes and Xie, Yonglong and Pierce, Andrew and Zaletel, Michael P and Yacoby, Amir and Vishwanath, Ashvin},
  journal={arXiv preprint arXiv:2112.13837},
  year={2021}
}

@article{polshyn2022topological,
  title={Topological charge density waves at half-integer filling of a moir{\'e} superlattice},
url={https://www.nature.com/articles/s41567-021-01418-6},
  author={Polshyn, Hryhoriy and Zhang, Yuxuan and Kumar, Manish A and Soejima, Tomohiro and Ledwith, Patrick and Watanabe, Kenji and Taniguchi, Takashi and Vishwanath, Ashvin and Zaletel, Michael P and Young, Andrea F},
  journal={Nature Physics},
  volume={18},
  number={1},
  pages={42--47},
  year={2022},
  publisher={Nature Publishing Group UK London}
}

@article{wilhelm2023non,
  title={Non-coplanar magnetism, topological density wave order and emergent symmetry at half-integer filling of moir{\'e} Chern bands},
url={https://www.scipost.org/10.21468/SciPostPhys.14.3.040?acad_field_slug=astronomy},
author={Wilhelm, Patrick and Lang, Thomas and Scheurer, Mathias S and L{\"a}uchli, Andreas},
  journal={SciPost Physics},
  volume={14},
  number={3},
  pages={040},
  year={2023}
}

@article{xie2023phase,
  title={Phase diagram of twisted bilayer graphene at filling factor $\nu$=$\pm$3},
url={https://journals.aps.org/prb/abstract/10.1103/PhysRevB.107.075156},
  author={Xie, Fang and Kang, Jian and Bernevig, B Andrei and Vafek, Oskar and Regnault, Nicolas},
  journal={Physical Review B},
  volume={107},
  number={7},
  pages={075156},
  year={2023},
  publisher={APS}
}

@article{PhysRevB.100.085109,
  title = {Magic angle hierarchy in twisted graphene multilayers},
  author = {Khalaf, Eslam and Kruchkov, Alex J. and Tarnopolsky, Grigory and Vishwanath, Ashvin},
  journal = {Phys. Rev. B},
  volume = {100},
  issue = {8},
  pages = {085109},
  numpages = {9},
  year = {2019},
  month = {Aug},
  publisher = {American Physical Society},
  doi = {10.1103/PhysRevB.100.085109},
  url = {https://link.aps.org/doi/10.1103/PhysRevB.100.085109}
}

@article{dong2023many,
  title={Many-body ground states from decomposition of ideal higher Chern bands: Applications to chirally twisted graphene multilayers},
url={https://journals.aps.org/prresearch/abstract/10.1103/PhysRevResearch.5.023166},
  author={Dong, Junkai and Ledwith, Patrick J and Khalaf, Eslam and Lee, Jong Yeon and Vishwanath, Ashvin},
  journal={Physical Review Research},
  volume={5},
  number={2},
  pages={023166},
  year={2023},
  publisher={APS}
}

\clearpage
\newpage
\onecolumngrid

\begin{appendix}
\section{HF Hamiltonian}\label{AppendixHF}
In this appendix, we provide details of our HF procedure.
We start with the normal-ordered Coulomb interaction \cite{parker2021field}
$$ H_V  =\frac{1}{2A}\sum_{\vec{q}} \sum_{\vec{k},\vec{k'}\in \textrm{mBZ}}V(\vec{q})[\Lambda_{-\vec{q}}(\vec{k})]_{\alpha \delta}[\Lambda_q(\vec{k}')]_{\beta \gamma}c^{\dagger}_{\vec{k},\alpha}c^{\dagger}_{\vec{k}',\beta}c_{\vec{k}'+\vec{q}, \gamma}c_{\vec{k}-\vec{q} ,\delta}, $$
where the momentum summation over $\vec{q}$ is not restricted to the mBZ. The form factor $\Lambda_{\vec{q}}(\vec{k})$ is defined as $[\Lambda_{\vec{q}}(\vec{k})]_{\alpha \beta} = \langle u_{\vec{k},\alpha}|u_{\vec{k}+\vec{q}, \beta} \rangle$ with flavor/band quantum numbers labeled by $\alpha, \beta$. By fixing the gauge to be periodic in mBZ such that $c^{\dagger}_{\vec{k}+\vec{G}} = c^{\dagger}_{\vec{k}}$, the form factor satisfies $\Lambda_{\vec{q}}(\vec{k}) = \Lambda_{\vec{q}}(\vec{k+G})$, with moir\'e reciprocal vector $\vec{G}$ \cite{bultinck2020ground}.

For the fermions, there are four possible contractions, $\langle c^\dagger_{\vec{k},\alpha} c_{\vec{k}'+\vec{q},\gamma}\rangle ,\, \langle c^\dagger_{\vec{k},\alpha} c_{\vec{k}-\vec{q},\delta}\rangle, \,\langle c^\dagger_{\vec{k}',\beta} c_{\vec{k}'+\vec{q},\gamma}\rangle, \, \langle c^\dagger_{\vec{k'},\beta} c_{\vec{k}-\vec{q},\delta}\rangle$. By imposing translation symmetry breaking ansatz with momentum vector $\vec{Q}$ as $\langle c^\dagger_{\vec{k},\alpha} c_{\vec{k'}+\vec{q},\gamma}\rangle = [P(\vec{k},\vec{Q})]_{\alpha \gamma} \delta_{\vec{k},\vec{k'}+\vec{q}+\vec{Q}}$ and $\langle c^\dagger_{\vec{k},\alpha} c_{\vec{k}-\vec{q},\delta}\rangle = [P(\vec{k},\vec{Q})]_{\alpha\delta} \delta_{\vec{k},\vec{k}-\vec{q}+\vec{Q}}$, we get the following Hartree and Fock Hamiltonian as $V_{\textrm{Hartree(Fock)}}(\vec{Q}) =\sum\limits_{\vec{k}\in \textrm{mBZ}}c^{\dagger}_{\vec{k}}h_{\textrm{Hartree(Fock)}}(\vec{k}, \vec{Q})c_{\vec{k}+ \vec{Q}}$, with
\begin{subequations}
\begin{align}
\begin{split}
h_{\textrm{Hartree}}(\vec{k},\vec{Q}) =\frac{1}{A}\sum_GV(\vec{G}+\vec{Q})\Lambda_{\vec{G}+\vec{Q}}(\vec{k})\sum_{\vec{k'} \in \textrm{BZ}} \textrm{Tr}(P(\vec{k'}, \vec{Q})\Lambda^*_{\vec{G}+\vec{Q}}(\vec{k'}- \vec{Q} -\vec{G})) \end{split} \\
\begin{split}h_{\textrm{Fock}}(\vec{k},\vec{Q}) =-\frac{1}{A}V_{\vec{q}}\Lambda_{\vec{q}}(\vec{k}) P^T([\vec{k} + \vec{Q}+\vec{q}], \vec{Q})\Lambda^{\dagger}_{\vec{q}}(\vec{k}+ \vec{Q}),
\end{split}
\end{align}
\end{subequations}
where $[\vec{k}] = \vec{k}-\vec{G}$ is the backfolded momentum into the mBZ with a corresponding moir\'e reciprocal vector $\vec{G}$. By defining $h(\vec{k}, \vec{Q}) \equiv h_{\textrm{Hartree}}(\vec{k}, \vec{Q})+h_{\textrm{Fock}}(\vec{k}, \vec{Q})$, we get the following relation: $h(\vec{k},\vec{Q}) = h^{\dagger}(\vec{k}+\vec{Q},\vec{Q})$ due to the periodic gauge.

\subsection{$1Q$ ansatz}
Let us first consider the $1Q$ ansatz, where the moir\'e translation symmetry is broken only along a single momentum of the three $M$ points in the mBZ. Since the three $M$ points are related by symmetry, we can set $\vec{Q}$ to be any single $M$ point without loss of generality. For $\mathcal{Q} = \{\Gamma,  \vec{Q}\}$, by defining the spinor $\mathbf{c}_{\mathcal{Q}\vec{k}} \equiv (c_{\vec{k}} \ c_{\vec{k}+\vec{Q}})^T$, we can write the full HF Hamiltonian, including the contributions from the non-interacting part $h_{0}(\vec{k})$  [in diagonalized form given by \equref{eq:BM} in the main text] as $H_{\textrm{HF}} = \sum\limits_{\vec{k}\in \textrm{mBZ}_{1Q}} \textbf{c}^{\dagger}_{\mathcal{Q}\vec{k}}\textbf{h}_{\textrm{HF}}(\vec{k}, \mathcal{Q})c_{\mathcal{Q}\vec{k}}$, where momentum $\vec{k}$ summation is restricted to the $\textrm{mBZ}_{1Q}$ and 
\begin{align}
\textbf{h}_{\textrm{HF}}(\vec{k}, \mathcal{Q})=\left ( \begin{array}{cccc} h_{0}(\vec{k})+h(\vec{k},0) & h(\vec{k},\vec{Q}) \\
              h^{\dagger}(\vec{k},\vec{Q})  & h_{0}(\vec{k}+\vec{Q})+h(\vec{k}+\vec{Q},0)
             \end{array} \right). 
\end{align}

\subsection{3$Q$ ansatz}
Now, we consider the $3Q$ ansatz, in which the moir\'e translation symmetry is broken along all three $M$ points in the mBZ. By denoting three $M$ points as $\vec{Q}_1, \vec{Q}_2, \vec{Q}_3$ and $\mathcal{Q} = \{\Gamma, \vec{Q}_1, \vec{Q}_2, \vec{Q}_3\}$, we define the spinor $\mathbf{c}_{\mathcal{Q}\vec{k}} \equiv (c_{\vec{k}} \ c_{\vec{k}+\vec{Q_1}}  \ c_{\vec{k}+\vec{Q_2}}  \ c_{\vec{k}+\vec{Q_3}})^T$, which allows the full HF Hamiltonian to be written as $H_{\textrm{HF}} = \sum\limits_{\vec{k}\in \textrm{mBZ}_{3Q}}\mathbf{c}^{\dagger}_{\mathcal{Q}\vec{k}}\mathbf{h}_{\textrm{HF}}(\vec{k}, \mathcal{Q}) \mathbf{c}_{\mathcal{Q}\vec{k}}$ with 
\begin{align}
\textbf{h}_{\textrm{HF}}(\vec{k}, \mathcal{Q})=\left ( \begin{array}{cccc} h_{0}(\vec{k})+h(\vec{k},0) & h(\vec{k},\vec{Q}_1) & h(\vec{k},\vec{Q}_2) & h(\vec{k},\vec{Q}_3) \\
              h^{\dagger}(\vec{k},\vec{Q}_1)  & h_{0}(\vec{k_1})+h(\vec{k_1},0) & h(\vec{k_1},\vec{Q}_3)  &h(\vec{k_1},\vec{Q}_2) \\
             h^{\dagger}(\vec{k},\vec{Q}_2) &h^{\dagger}(\vec{k_1},\vec{Q}_3)&h_{0}(\vec{k_2})+h(\vec{k_2},0)&h(\vec{k_2}, \vec{Q}_1)\\
              h^{\dagger}(\vec{k},\vec{Q}_3)&   h^{\dagger}(\vec{k_1},\vec{Q}_2)&h^{\dagger}(\vec{k_2},\vec{Q}_1)&h_{0}(\vec{k_3})+h(\vec{k_3},0)
             \end{array} \right),
\end{align}
and $\vec{k}_i \equiv \vec{k}+\vec{Q}_i$. Here, the momentum summation over $\vec{k}$ is also restricted to the appropriately adjusted $\textrm{mBZ}_{3Q}$.

\newpage
\subsection{Folded Brillouin zone}\label{FoldedBZ}
In this section, we discuss the folded Brillouin zones (mBZ$_{\mathcal{Q}}$) for the numerical implementation. We first choose the momentum grid set for the original mBZ as $\textrm{mBZ} = \{\frac{n}{N}\vec{G}_1+\frac{m}{N}\vec{G}_2| \ 0 \leq n, m \leq N-1 \}$, where we set $\vec{G}_1 = K_{\theta}(\sqrt{3},0), \ \vec{G}_2 = K_{\theta}(-\frac{1}{2},-\frac{\sqrt{3}}{2})$ with $K_{\theta} = 8\pi \ \textrm{sin}(\theta/2)/{3a_0}$, twist angle $\theta$ and graphene lattice constant $a_0$. Then, we define the $\textrm{mBZ}_{1Q} \ (\mathcal{Q}=\{\Gamma,M_1\})$ and $\textrm{mBZ}_{3Q} \ (\mathcal{Q}=\{\Gamma,M_1, M_2, M_3\})$ for the $1Q$ and $3Q$ ans\"atze respectively as
$$\textrm{mBZ}_{1Q} = \{\frac{n}{N}\vec{G}_1+\frac{m}{N}\vec{G}_2| \ 0 \leq n \leq \frac{N}{2}- 1, \ 0 \leq m \leq N- 1 \}$$
$$\textrm{mBZ}_{3Q} = \{\frac{n}{N}\vec{G}_1+\frac{m}{N}\vec{G}_2| \ 0 \leq n \leq \frac{N}{2}- 1, \ 0 \leq m \leq \frac{N}{2}- 1 \}$$
Numerically, we choose $N$ to be an even number to make $N/2$ an integer and also to be a multiple of 3 to ensure that the chosen momentum grid includes the $K$ point in the mBZ. 

\section{Additional data of spin chirality and product}\label{OtherSublatticesLayers}
\begin{figure*}[t]
	\centering
	\includegraphics[width=1\columnwidth]{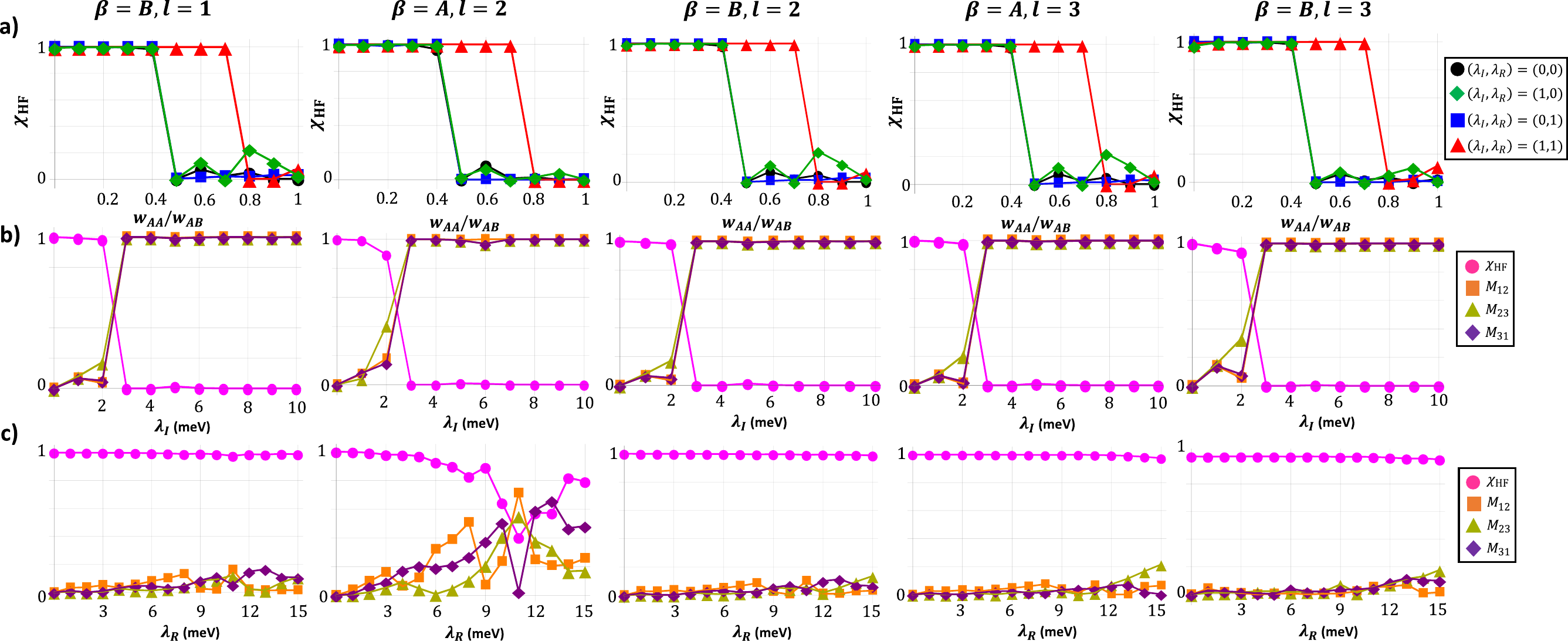}
	\caption{{\bf Spin chirality and inner product of spin magnetization for $3Q$ states with different choices of sublattice and layer degrees of freedom} 
    \justifying
    (\textbf{a}): Spin chirality $\chi_{\textrm{HF}}$ as a function of $w_{AA}/w_{AB}$ with different SOC parameters and sublattice $\beta$, layer $l$ degrees of freedom. The SOC parameters are in units of meV.  (\textbf{b}): Spin chirality $\chi_{\textrm{HF}}$ and inner products $M_{12}, M_{23}, M_{31}$ as a function of $\lambda_I$ for fixed $w_{AA}/w_{AB}=0, \lambda_R=0$ with different sublattice, layer degrees of freedom. (\textbf{c}): Spin chirality $\chi_{\textrm{HF}}$ and inner products $M_{12}, M_{23}, M_{31}$ as a function of $\lambda_R$ for fixed $w_{AA}/w_{AB}=0, \lambda_I=0$ with different sublattice, layer degrees of freedom. The system size is $N=12\times12$.}
	\label{Fig14:layerlatticeextradata}
    \end{figure*}

In this section, we show the additional spin chirality $\chi_{\textrm{HF}}$ and product $M_{ij}$ data for different choices of sublattice $\beta$ and layer $l$ degrees of freedom based on Eq.\ (\ref{eq:chirality}), see \figref{Fig14:layerlatticeextradata}. Figure~\ref{Fig14:layerlatticeextradata}(a) and (b) show that for small SOC strength ($= 1 \ \textrm{meV}$), or the presence of Ising SOC only, the qualitative features of $\chi_{\textrm{HF}}$ and $M_{ij}$ do not change for different choices of sublattice and layer degrees of freedom. In contrast, Fig.\ \ref{Fig14:layerlatticeextradata}(c) shows that the presence of large Rashba SOC makes $\chi_{\textrm{HF}}$ and $M_{ij}$ much more sensitive to the choice of sublattice and layer degrees of freedom. We expect this behavior to arise, due to the sublattice dependent term of Rashba SOC Hamiltonian (see Eq.\ (\ref{eq:SOCham})). 

\section{Including the valence bands}\label{WithValenceBands}

\begin{figure*}[t]
	\centering
	\includegraphics[width=0.8\columnwidth]{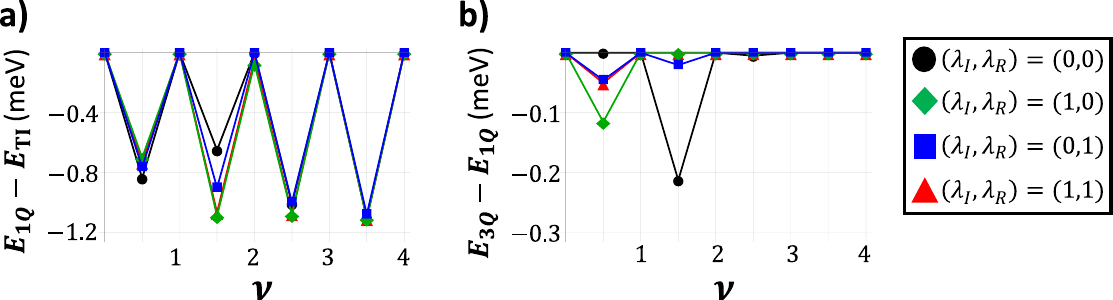}
	\caption{{\bf Two-bands HF data} Filling-factor dependence of two-bands HF energy difference per moir\'e unit cell
    \justifying
    (\textbf{a}): between the solution obtained from TI and $1Q$ ans\"atze and (\textbf{b}): $1Q$ and $3Q$ ans\"atze with different SOC parameters. The SOC parameters are in units of meV. The system size is $N=12\times12$. }
	\label{Fig13:twobandenerg}
    \end{figure*}
    
Finally, we extend our previous HF calculations, by including the valence bands (i.e., one more band per spin and valley).
This involves projecting the interacting Hamiltonian onto 
both conduction and valence bands, and perform an analogous HF calculation. As mentioned in Sec.~\ref{sec:HF}, compared to the HF calculation with interaction projected onto the conduction bands only, including valence bands additionally gives information about correlated states at charge neutrality ($\nu=0$), where the valence bands are fully occupied. 
Here, we point out that hybridization between conduction and valence bands will be facilitated by the small single-particle band gap of SOC-TMBG ($\sim 1 \ \textrm {meV}$), compared to the Coulomb interaction energy scale ($\sim 100\ \textrm{meV}$). 

\subsection{Competition between TI, $1Q$ and $3Q$ states}
Similar to \secref{EnCompTI1Q} and \secref{SBH1Q}, we first compare the two-bands HF energy per moir\'e unit cell for the HF solutions obtained from TI, $1Q$ and $3Q$ ans\"atze with different filling-factors and SOC parameters. The results are shown in Fig.\ \ref{Fig13:twobandenerg}(a) and Fig.\ \ref{Fig13:twobandenerg}(b). Compared to single-band HF, we first clearly see that regardless of SOC, a $1Q$ state is energetically favored over a TI state for all half-integer fillings, where a finite SOC only quantitatively affects the energy difference between $1Q$ and TI states. For the integer fillings, we see that the HF solutions of the $1Q$ ansatz also converge to TI states, except at $(\lambda_I, \lambda_R) = (1,0), \ (1,1) \ \textrm{meV}$ and $\nu=2$, where a $1Q$ state is energetically favored over a TI state. In detail, for $\nu=2$, our HF calculation determines the energy difference per moir\'e unit cell between $1Q$ and TI states to be 0.07 meV for $(\lambda_I, \lambda_R) = (1,0) \ \textrm{meV}$ and 0.04 meV for $(\lambda_I, \lambda_R) = (1,1) \ \textrm{meV}$. We do not take this as a contradiction to the single-band HF data because we cannot conclude that a change in the ground state occurs based on such a small numerical energy difference.

In addition, we see that for half-integer fillings, the HF solutions with $3Q$ ansatz converge to either $3Q$ or $1Q$ states depending on SOC parameters. For example, our HF data show that at $\nu = 1/2$, the HF solution with $3Q$ ansatz converges to a $1Q$ state when SOC is absent, while it converges to a $3Q$ state when SOC is present. However, for all integer fillings, we see that the HF solution with $3Q$ ansatz converges to a TI state, also except at $(\lambda_I, \lambda_R ) = (1,0), \ (1,1) \ \textrm{meV}$ and $\nu=2$, where it converges to a $1Q$ state.

\subsection{Band gap and band hybridization order parameter of HF ground state}\label{BandGap}
\begin{figure*}[t]
	\centering
	\includegraphics[width= 1\columnwidth]{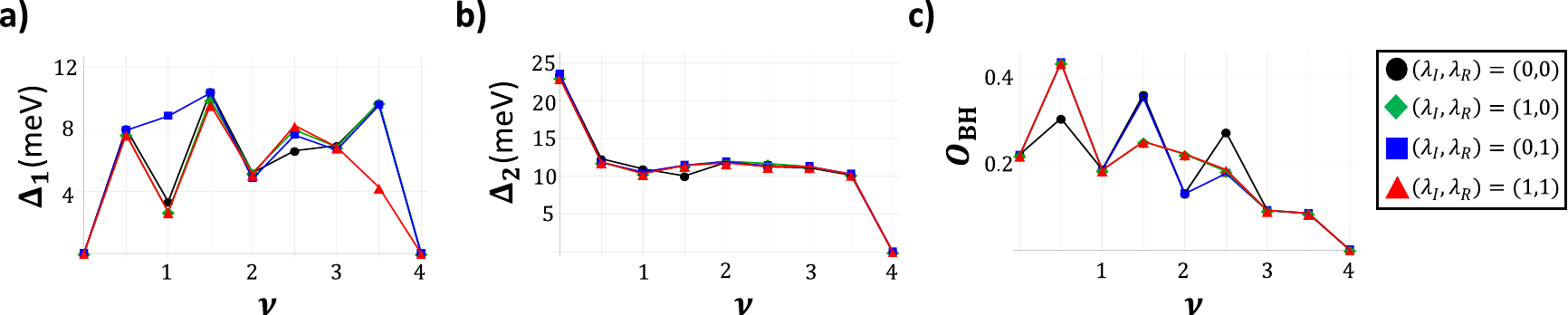}
	\caption{{\bf Band gap and band hybridization order parameter of HF ground state} 
    \justifying
    Filling-factor dependence of indirect energy band gap for HF ground state obtained from (\textbf{a}) single band HF  (\textbf{b}) two band HF with different SOC parameters. The SOC parameters are in units of meV. (\textbf{c}) Filling factor dependence of band hybridization order parameter $O_{\textrm{BH}}$ with different SOC parameters. The system size $N = 12 \times12$. 
	}
	\label{FigS1:HFband}
\end{figure*}

In this section, we show the filling-factor dependence of the indirect band gap $\Delta_1$ and $\Delta_2$ for the HF ground states obtained from a single-band HF and two-bands HF in figure \ref{FigS1:HFband}(a) and (b), respectively, with different SOC parameters. For nonzero half-integer fillings, we see that the band gap of a HF ground state obtained from a two-bands HF calculation becomes larger than that from a single-band HF. We expect this to happen because of the finite band hybridization.

To see this, we plot the band hybridization order parameter $O_{\textrm{BH}}$ of the HF ground state obtained from a two-bands HF calculation in Fig.\ \ref{FigS1:HFband}(c). For given ground state $\textrm{GS}\in\{\textrm{TI}, 1Q, 3Q\}$ with its correlator $\vec{P}_{\textrm{GS}}(\vec{k})$, momentum $\vec{k}\in\textrm{mBZ}_{\textrm{GS}}$ and band indices $b,b'$ (other spin, valley and additional indices from the translation symmetry breaking are omitted for simplicity), we define the band-off diagonal matrix $\vec{P}'_{\textrm{GS}}(\vec{k}) \equiv \vec{P}_{\textrm{GS}}(\vec{k})(\mathbbm{I}-\delta_{b,b'})$ with identity matrix $\mathbbm{I}$. The band hybridization order parameter $O_{\textrm{BH}}$ can be then defined as the Frobeniusnorm of $\vec{P}'_{\textrm{GS}}(\vec{k})$
\begin{align}
O_{\textrm{BH}} = \frac{1}{N_{\textrm{GS}}}\sum_{\vec{k}\in\textrm{mBZ}_{\textrm{GS}}} \sqrt{\vec{P}'^{\dagger}_{\textrm{GS}}(\vec{k})\vec{P}'_{\textrm{GS}}(\vec{k})},
\end{align}
where $N_{\textrm{GS}}$ is the system size of momentum grid corresponding to the HF ground state. We clearly see the nonzero $O_{\textrm{BH}}$ for all considered integer and half-integer fillings, which will lead to enhancement of a band gap compared to our single-band HF band structure. 

\end{appendix}

\end{document}